%% file: ms.tex
\documentclass[a4paper,11pt]{article}

\usepackage{jcappub}
\usepackage[T1]{fontenc}
\usepackage{pdfpages}
\usepackage[numbers,sort&compress]{natbib}
\usepackage{xcolor}
\usepackage{ulem}

\newcommand{\specialcell}[2][c]{%
  \begin{tabular}[#1]{@{}c@{}}#2\end{tabular}}

\def\LCDM{$\Lambda$CDM~}

\input{derivation_commands}
\title{The full Boltzmann hierarchy for dark matter-massive neutrino interactions}

\author[a]{Markus R. Mosbech,}
\author[a]{Celine Boehm,}
\author[b]{Steen Hannestad,}
\author[c]{Olga Mena,}
\author[d]{Julia Stadler,}
\author[e]{and Yvonne Y. Y. Wong}

\emailAdd{mmos6302@uni.sydney.edu.au}
\emailAdd{celine.boehm@sydney.edu.au}
\emailAdd{sth@phys.au.dk}
\emailAdd{omena@ific.uv.es}
\emailAdd{jstadler@mpe.mpg.de}
\emailAdd{yvonne.y.wong@unsw.edu.au}

\affiliation[a]{School of Physics, University of Sydney, Camperdown, NSW 2006, Australia\\
Sydney Consortium for Particle Physics and Cosmology}
\affiliation[b]{Department of Physics and Astronomy, Aarhus University, DK-8000 Aarhus C, Denmark}
\affiliation[c]{IFIC, Universidad de Valencia-CSIC, 46071, Valencia, Spain}
\affiliation[d]{Max Planck Institute for Extraterrestrial Physics, Giessenbachstraße 1, 85748 Garching, Germany}
\affiliation[e]{School of Physics, The University of New South Wales, Sydney NSW 2052, Australia,\\
Sydney Consortium for Particle Physics and Cosmology}

\abstract{

The impact of dark matter-neutrino interactions on the measurement of the cosmological parameters has been investigated in the past in the context of massless neutrinos exclusively. Here we revisit the role of a neutrino-dark matter coupling in light of ongoing cosmological tensions  by implementing the full Boltzmann hierarchy for three massive neutrinos. Our tightest 95\% CL upper limit on the strength of the interactions, parameterized via $u_\chi =\frac{\sigma_0}{\sigma_{Th}}\left(\frac{m_\chi}{100 \text{GeV}}\right)^{-1}$, is $u_\chi\leq3.34 \cdot 10^{-4}$, arising from a combination of Planck TTTEEE data, Planck lensing data and SDSS BAO data. This upper bound is, as expected, slightly higher than previous results for interacting massless neutrinos, due to the correction factor associated with neutrino masses. We find that these interactions significantly relax the lower bounds on the value of $\sigma_8$ that is inferred in the context of \LCDM from the Planck data, leading to agreement within 1-2$\sigma$ with weak lensing estimates of $\sigma_8$, as those from KiDS-1000. However, the presence of these interactions barely affects the value of the Hubble constant $H_0$.
}

\begin{document}
\begin{flushright}
	{\large \tt CPPC-2020-17}
\end{flushright}
\maketitle

\section{Introduction}

The cosmological \LCDM model is tremendously successful in describing scales ranging from the cosmological volume to the sizes of galaxies. In its most basic form, it depends on six free parameters only, five of which are determined to better than $1\%$ accuracy to date \cite{Planck2018I}. In particular Cosmic Microwave Background data (CMB), together with measurements of the universe's Large Scale Structure (LSS), allow for such precise constraints. Present data is well described by the minimal parameterization of the \LCDM model, which treats neutrinos either as massless particles or sets their masses to the minimum value allowed by neutrino oscillations, $m_\nu \sim 0.06\,\mathrm{eV}$ \cite{Esteban:2018azc}. Cosmology already imposes the tightest limits available on the neutrino mass scale: $m_\nu \leq 0.241\,\mathrm{eV}$ from the CMB alone and $m_\nu\leq 0.120\,\mathrm{eV}$ from a joint analysis of CMB and Baryonic Acoustic Oscillations (BAO) \cite{Planck2018I}, both at $95\%$~CL. These constraints surpass not only the current sensitivity of kinematic terrestrial neutrino-mass experiments, such as KATRIN \cite{Osipowicz:2001sq} (which currently reports a $95\%$~CL upper limit of $m_\nu \leq 1.1\,\mathrm{eV}$ \cite{Aker:2019uuj}), but also future prospects, which plan to improve the present limit to $0.2\,\mathrm{eV}$. Planned LSS surveys \cite{Aghamousa:2016zmz, 2011arXiv1110.3193L, Santos:2015gra} will further increase the sensitivity to $\sim 0.02\,\mathrm{eV}$ and thus provide hope for a cosmological detection of neutrino masses \cite{Font-Ribera:2013rwa,Brinckmann:2018owf}.

Constraints on the neutrino mass from CMB and LSS, however, depend on the cosmological scenario considered, and alternatives to the minimal \LCDM scenario are attractive for a number of reasons. \LCDM faces challenges in effectively describing structures on sub-galactic scales \cite{Bullock:2017xww, Klypin:1999uc, Moore:1999nt, Flores:1994gz, Moore:1994yx, BoylanKolchin:2011de} as well as tensions between different experiments. Most notably, there is a disagreement at more than $4\sigma$ \cite{Verde:2019ivm,Valentino2020H0} between the value of $H_0$ determined from early \cite{Planck2018,DESY1H0,ACTDR4,6dFH0,SDSSDR7clustering,SDSSDR12clustering} and late-type \cite{Riess2019ApJ,H0licowXIII,CCHPVIII,Miras} cosmological probes, with the latter favouring higher values. Another tension is the discrepancy in $\sigma_8/S_8$ clustering parameter between Planck CMB data and other measurements from weak lensing surveys~\cite{DES-1-galaxy-wl,DES-1-shear,KiDS-1000}. Apart from new physics, instrumental effects have been brought forward to explain the discrepancy \cite{LockdownPerspective}. Finally, the \LCDM model remains a phenomenological description where the nature of two of its principal components, dark matter (DM) and dark energy, are still unknown.

Complementary to collider searches and direct detection efforts, the microscopic properties of DM can be investigated from cosmological observations, by relaxing the assumptions of a cold, collisionless single fluid. Models of warm DM \cite{Dodelson:1993je,Bode:2000gq,Asaka:2005an,Viel:2005qj,Boyarsky:2009ix,Viel:2013fqw,Abazajian:2001nj,Boyarsky:2008xj,Dolgov:2000ew}, self-interacting DM \cite{Carlson:1992fn, deLaix:1995vi, Spergel:1999mh, Dave:2000ar, Boehm:2000gq, Boehm:2004th, Creasey:2016jaq, Rocha:2012jg, Kim:2016ujt, Huo:2017vef, Markevitch:2003at, Randall:2007ph} and DM interactions with baryons \cite{Boehm:2000gq, Boehm:2004th,Chen:2002yh, Dvorkin:2013cea, Dolgov:2013una, CyrRacine:2012fz, Prinz:1998ua}, photons \cite{Boehm:2000gq, Boehm2001InteractingMatter, Boehm:2001hm, Boehm:2004th, Sigurdson:2004zp, CyrRacine:2012fz, Dolgov:2013una,  Wilkinson:2013kia, Boehm:2014vja, Schewtschenko:2014fca, Schewtschenko:2015rno, Escudero:2015yka, McDermott:2010pa, Diacoumis:2017hff, Ali-Haimoud:2015pwa, Stadler:2018jin}, neutrinos \cite{Boehm:2000gq, Boehm:2004th, Mangano:2006mp, 2010PhRvD..81d3507S, Wilkinson:2014ksa, DiValentino:2017oaw, Ali-Haimoud:2015pwa} or dark radiation \cite{Das:2017nub, Kaplan:2009de, Diamanti:2012tg,  Buen-Abad:2015ova, Lesgourgues:2015wza, Ko:2017uyb, Escudero:2018thh, Das:2010ts, Archidiacono:2019wdp} have therefore received plenty of attention in the literature. Recently, the case of DM simultaneously interacting with multiple other species has also been considered \cite{Becker:2020hzj}. In the context of DM-neutrino interactions, however, only the massless limit has been implemented, making it e.g. impossible to simultaneously derive constraints on the interaction strength of three interacting neutrinos and their mass scale. To the best knowledge of the authors, no previous study has re-derived the full Boltzmann hierarchy for massive neutrinos in the presence of interactions with dark matter.

In light of the imminent improvements on the neutrino mass scale from cosmological observations, we consider it timely to update the formalism for DM-$\nu$ interactions. Here, we present the full Boltzmann hierarchy for massive neutrinos interacting with DM through a constant scattering term (similar to e.g. Thomson scattering) and its implementation in the Boltzmann solver CLASS \cite{CLASSII}. The modified CLASS version developed for this paper can be used to create initial conditions for N-body simulations, which study the effect of DM-$\nu$ interactions on smaller scales\footnote{Our code is publicly available under \url{https://github.com/MarkMos/CLASS_nu-DM}}. We use our results to update constraints on the neutrino interaction rate from cosmological observations through a MCMC approach using the publicly available code MontePython \cite{Brinckmann:2018cvx,Audren:2012wb}. For a minimal neutrino mass scale of $0.06\,\mathrm{eV}$, we find that the correct treatment of the neutrino mass slightly loosens the bounds on the cross-section, due to the appearance of an momentum-dependent factor $p^2/E^2$ in the scattering term. 

The structure of the paper is as follows. In Section \ref{sec:boltzmanneq} we review the linearized relativistic Boltzmann equations for $\Lambda$CDM, then we present the modification from DM-$\nu$ interactions in Section \ref{sec:modboltzmanneq}. The impact on the CMB and on the matter power spectrum are discussed in Section \ref{sec:conclusion}, together with the parameter constraints obtained from current cosmological observations. The detailed derivation of the modified Boltzmann equations and the accuracy of its numerical implementation are presented in Appendix \ref{sec:derivation}) and  Appendix \ref{sec:fluidapprox}), respectively. We shall also illustrate the results for all the cosmological parameters considered in this work in Appendix \ref{sec:2dplots}.

\section{The Boltzmann equation}
\label{sec:boltzmanneq}
For a large fraction of its evolution history, the universe is well-described by the linearised Boltzmann equations. Beginning in a very homogeneous state with only minor deviations, the linear Boltzmann treatment works extremely well, until the onset of nonlinear structure formation in the late universe. 

The basis of the framework is the relativistic Boltzmann equation, as presented in e.g. references~\cite{MaBertschinger,Oldengott_2015}
\begin{equation}
    P^\alpha \frac{\partial f}{\partial x^\alpha}-\Gamma^\gamma_{\alpha\beta}P^\alpha P^\beta \frac{\partial f}{\partial P^\gamma} = m\left(\frac{\partial f}{\partial \tau}\right)_C\,,
\end{equation}
which describes the evolution of the distribution function $f$ for a given species with mass $m$, defined through the number of particles in a phase-space volume element 
\begin{equation}
    f(\mathbf{x},\mathbf{P},\tau)dx^1 dx^2 dx^3 dP^1 dP^2 dP^3 = dN \,,
\end{equation}
where $x^i$ are the comoving coordinates and $P^i$ their canonical momentum conjugate. We can then define the perturbation to the distribution function, $\Psi$, using the comoving momentum $\mathbf{p},$
\begin{equation}
   f(\mathbf{x},\mathbf{p},\tau) = f_0 (p) \left[1+\Psi(\mathbf{x},\mathbf{p},\tau )\right].
\end{equation}
The Boltzmann equation retains the gauge freedom of general relativity, so it is necessary to solve it in a specific gauge. The Boltzmann equation for $\Psi$, reads, in the Newtonian gauge, as
\begin{equation}
	\frac{\partial\Psi}{\partial\tau} + i \frac{p}{E}\left(\mathbf{k}\cdot\hat{\mathbf{n}}\right)\Psi + \frac{d\ln\fz{}}{d\ln p} \left[\dot{\phi}-i\frac{E}{p}\left(\hat{\mathbf{k}}\cdot\hat{\mathbf{n}}\right)\psi\right] = \frac{1}{f_0} \left(\frac{\partial f}{\partial \tau}\right)_C \,,
\end{equation}
where $\phi$ and $\psi$ are the gauge fields of the Newtonian gauge, $E$ and $p$ are particle energy and comoving momentum and $\hat{\mathbf{n}}$ is the unit vector pointing in the direction of the momentum. 

For most standard species, it is possible to perform a Legendre decomposition and integrate analytically over the particle momentum $p$ and then evolve the integrated quantities (e.g. $\delta$, $\theta$).  However such a simplification is not possible for  for massive neutrinos. The fact that they have a small but nonzero mass, means they are not inherently ultrarelativistic (like photons) or nonrelativistic (like baryons or cold dark matter (CDM)) and prevents the use of various approximations. Instead, it is required to solve the hierarchy in a momentum-dependent way. In practice, this is done by separately solving the hierarchy for a set of momentum bins, as described in e.g. \cite{MaBertschinger,Lesgourgues2011TheRelics}.

The Boltzmann equations for conventional cold dark matter are very simple, since it only interacts gravitationally. 
In the Newtonian gauge, the dark matter fluid is described by the equations
\begin{subequations}
\begin{align}
    \dot{\delta}_{\text{cdm}} &= -\theta_\text{cdm} + 3\dot{\phi}~,\\
    \dot{\theta}_\text{cdm} &= -\frac{\dot{a}}{a}\theta_\text{cdm} + k^2 \psi~.
\end{align}
\end{subequations}
The non-interacting Boltzmann hierarchy for massive neutrinos in the Conformal Newtonian gauge (using the conventions of reference~\cite{MaBertschinger}) are:
\begin{subequations}
\begin{align}
    \frac{\partial \Psi_0}{\partial \tau} &=-\frac{pk}{\Ep}\Psi_1 - \dot{\phi} \frac{d\ln\fz{}}{d\ln p}~,\\
    \frac{\partial \Psi_1}{\partial \tau} &=\frac{1}{3}\frac{pk}{\Ep} \left(\Psi_0-2\Psi_2\right) - \frac{\Ep k}{3p}\psi\frac{d\ln\fz{}}{d\ln p} ~,\\
    \frac{\partial\Psi_l}{\partial\tau} &=
    \frac{1}{2l+1}\frac{pk}{\Ep} 
    \left(l\Psi_{l-1}-(l+1)\Psi_{l+1}\right), \quad l\geq2~.
\end{align}
\end{subequations}

\section{Modified Boltzmann equations in presence of neutrino-dark matter interactions}
\label{sec:modboltzmanneq}
We introduce a new interaction between massive neutrinos and a dark matter species similarly to the case  of  massless neutrinos or photons interacting with  dark matter ~\cite{Wilkinson2014ConstrainingStructure,DiValentino2018ReducingInteractions,Stadler2019FirstDamping,Stadler2020IsMasses}. 

We shall assume a constant cross-section scattering between the massive neutrinos ($\nu$) and the DM particles ($\chi$), where the total neutrino energy is much smaller than the DM mass. This corresponds to a mass-degeneracy between the mediator and the dark matter particle \cite{Boehm:2000gq,Campo:2017nwh}. Solving the scattering integral for a massive neutrino with momentum $p$:
\begin{multline}
	C(p) = \frac{1}{E_\nu \! \left(\mathbf{p}\right)}\int \intdnutwo{\mathbf{p}'} \intdctwo{\mathbf{q}} \intdctwo{\mathbf{q}'} \left(2\pi\right)^4 \left|M\right|^2 \\
	\times\delta^4 \! \left(q + p - q' - p'\right) \left[\gq{'}\fp{'}\left(1-\fp{}\right)-\gq{}\fp{}\left(1-\fp{'}\right)\right],
\end{multline}
yields the collision term
\begin{equation}
C(p) = \frac{\ \sigma_0 n_\chi p^2}{E_\nu^2 \! \left(p\right)}\left[ 
f^{\left(1\right)}_0 \!\left(p\right) 
+ \frac{1}{2}f^{\left(1\right)}_2 \!\left(p\right) P_2 \! \left(\mu\right) 
-f^{\left(1\right)} \!\left(p,\mu\right) 
 - v_\chi \mu E_\nu \! \left(p\right) \frac{d\fz{}}{d p} \right],
\end{equation}
from which we can define a momentum-dependent interaction rate
\begin{equation}
	C_\chi\!(p) \equiv \frac{a \sigma_0 n_\chi p^2}{E_\nu^2 \! \left(p\right)}~.
\end{equation}
The extra factor of $a$ originates from the use of comoving coordinates and conformal time.
The cross section can be re-parameterized using 
\begin{equation}
    u_{\nu\chi} = \frac{\sigma_0}{\sigma_{\text{Th}}}\left(\frac{m_\chi}{100\, \text{GeV}}\right)^{-1},
\end{equation}
the interaction rate is then given by
\begin{equation}
\label{equation:chi}
	C_\chi = a \, u_{\nu\chi} \, \frac{\sigma_{\text{Th}} \rho_\chi}{100\, \text{GeV}}\frac{p^2}{E_\nu^2}~.
\end{equation}
The modifications to the Boltzmann hierarchy for massive neutrinos are gauge independent and consist of adding a collision term to the Euler equation and a damping term to the higher moments:
\begin{subequations}
\begin{align}
	\frac{\partial \Psi_1}{\partial \tau} &= \left[...\right] - C_\chi\frac{v_\chi E_\nu \! \left(p\right)}{3\fz{}} \frac{d\fz{}}{d p} - C_\chi\Psi_1~,\\
	\frac{\partial\Psi_2}{\partial\tau} &= \left[...\right] - \frac{9}{10}C_\chi\Psi_2~,\\
	\frac{\partial\Psi_l}{\partial\tau} &=
	\left[...\right]-C_\chi\Psi_l, \quad l\geq3~,
\end{align}
\end{subequations}
where $\left[...\right]$ represents the standard terms of the non-interacting case, omitted for clarity. Note that the $l=0$ equation is unmodified, as the interaction does not transfer energy at linear order.

The corresponding Boltzmann equations for DM are usually presented after being integrated over momentum. An integrated term has to be added to the $\theta_\chi$-equation,
\begin{equation}
\label{eq:dmterm}
\begin{split}
	\dot{\theta}_\chi &= \left[...\right] + K_\chi \, \frac{3}{4} k \, \frac{\int p^2 dp \, p \fz{} C_\chi(p)\left(\frac{\theta_\chi E_\nu \! \left(p\right)}{3k\fz{}} \frac{d\fz{}}{d p} + \Psi_1\right)}{\int p^2 dp\, p \fz{}}~,\\
	&= \left[...\right] + K_\chi \dot{\mu}_\chi \left(\theta_\nu - \theta_\chi\right)~,
\end{split}
\end{equation}
with $\left[...\right]$ representing the terms of the standard non-interacting DM equation, $\dot{\mu}_\nu$ is the interaction rate in terms of a $\nu$-DM opacity, as in e.g. \cite{Wilkinson2014ConstrainingStructure}. Since the integral in equation \ref{eq:dmterm} cannot be solved analytically, we do not offer an explicit definition of $\dot{\mu}_\nu$, but merely present it to serve as an interpretation of the integral. $K_\chi$ is the momentum conservation factor, given by
\begin{equation}
	K_\chi \equiv \frac{\rho_\nu + P_\nu}{\rho_\chi} = \frac{(1+w_\nu)\rho_\nu}{\rho_\chi}~,
\end{equation}
It is similar to the $R=\frac{4\rho_\gamma}{3\rho_b}$ found in the Thomson scattering term for baryons but not identical due to the fact that $w_\nu$ is not always $1/3$. Note that in our numerical code, $v_\chi$ can be expressed as $v_\chi = \theta_\chi/k$. For a detailed derivation of the interaction term and the Boltzmann hierarchy, see Appendix~\ref{sec:derivation}. It is also worth noting that CLASS does not always solve the full momentum-dependent hierarchy. In some regimes a fluid approximation is used instead, which greatly lowers computation time. In Appendix \ref{sec:fluidapprox} we describe how we consistently treat the momentum-dependent interaction rate $C_\chi (p)$ in this regime.

In our numerical implementation, the interacting species $\chi$ is included as a new species, independently of the non-interacting CDM component, and therefore allowing for only a fraction of the dark matter to interact with neutrinos. However, all of the results presented in this work are based on a scenario with 100\% of the dark matter being the interacting species $\chi$. Additionally, our framework allows for different cross sections for the different neutrino species, albeit the results presented here were obtained for three massive neutrino species with identical coupling to $\chi$. While in our publicly available modified CLASS version we have only implemented the interaction in the Newtonian gauge, the inclusion in the synchronous gauge would be straightforward.

\section{Numerical results}
\label{sec:numres}
\begin{figure}
	\centering
	\includegraphics[width=0.8\linewidth]{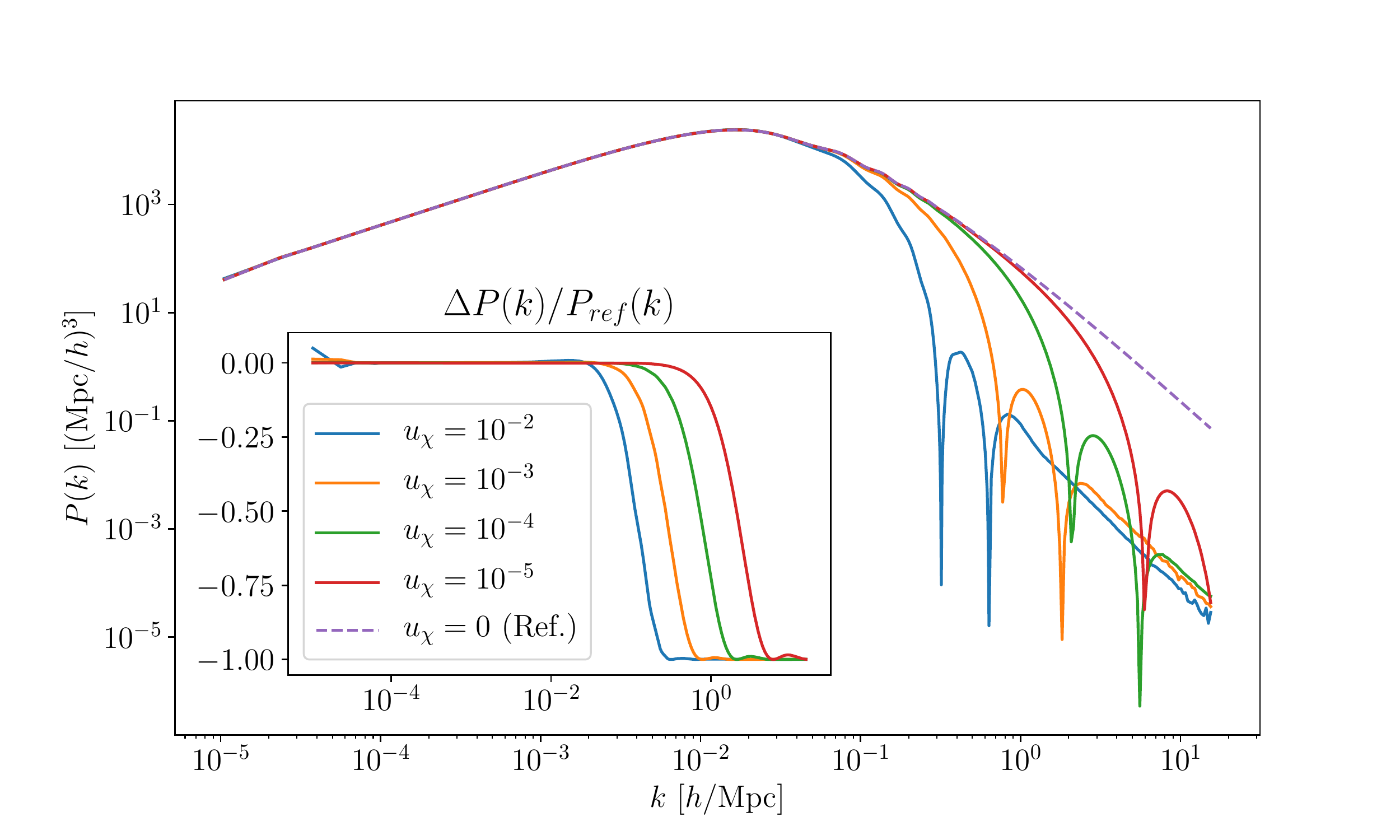}
	\caption{Matter power spectrum for different values of $u_\chi$ in a fiducial cosmology using Planck 2018 parameters \cite{Planck2018}, the insert panel the shows relative difference, compared to the non-interacting case.}
	\label{fig:Pk+Delta_ucomp}
\end{figure}

The largest impact of the interaction is on the matter power spectrum, as illustrated in figure~ \ref{fig:Pk+Delta_ucomp}, where the matter power spectrum is depicted for a range of $u_\chi$-values. The main effect is a damping on small scales, with the scale of the threshold for damping depending on $u_\chi$. The shape of the damped power spectrum is qualitatively similar to the results for massless neutrinos, see e.g.~\cite{Stadler2019How.,Wilkinson2014ConstrainingStructure,Mangano2006}. 

\begin{figure}
	\centering
	\includegraphics[width=0.8\linewidth]{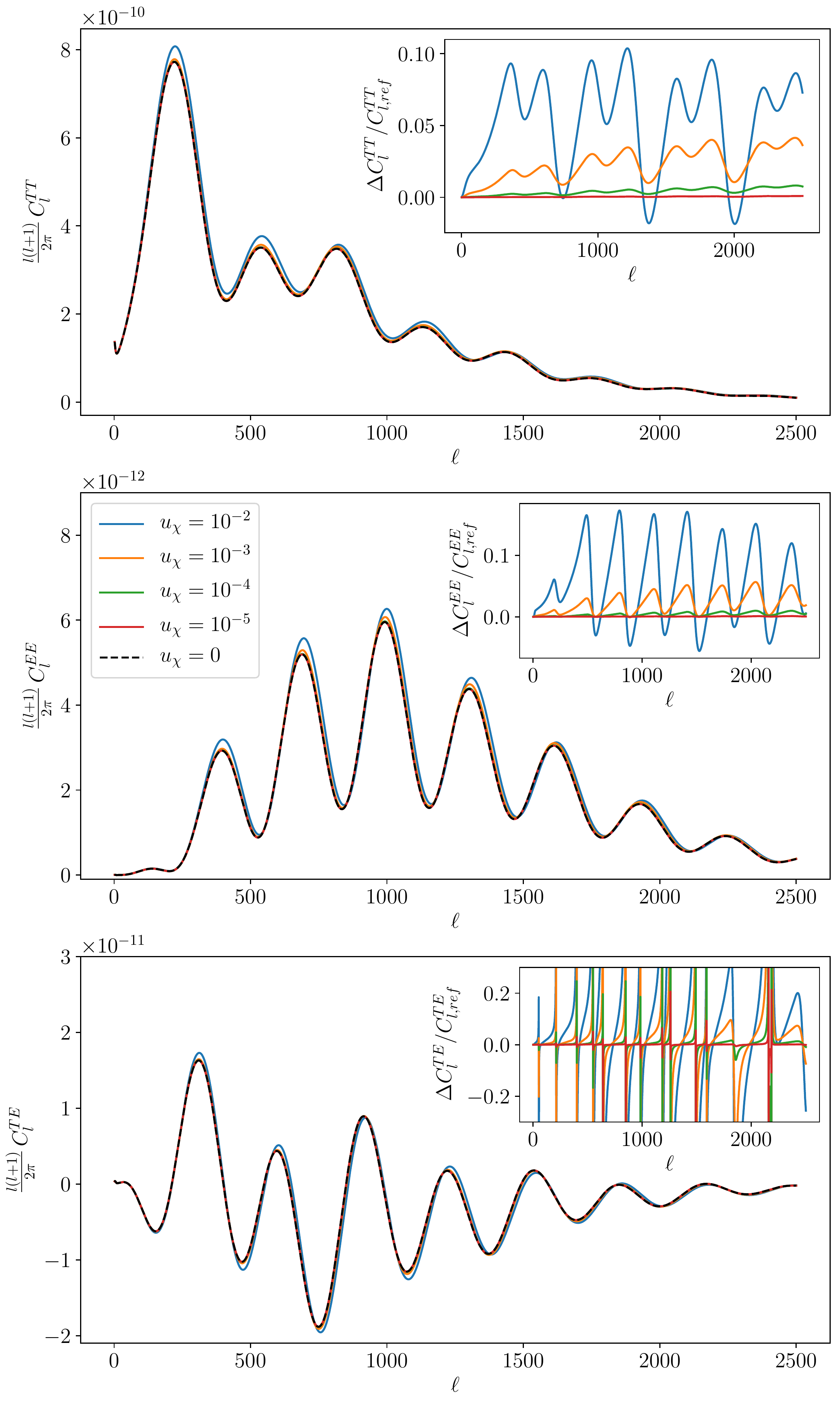}
	\caption{CMB TT, EE and TE power spectra for different values of $u_\chi$ in a fiducial cosmology using Planck 2018 parameters \cite{Planck2018}. The insert panels show the relative differences compared to the non-interacting case.}
	\label{fig:allCl}
\end{figure}

The effect on the CMB is illustrated in figure~\ref{fig:allCl}, see also references~\cite{DiValentino2018ReducingInteractions,Wilkinson2014ConstrainingStructure,Mangano2006}. As carefully  explained in  \cite{Cyr_Racine_2014,Wilkinson2014ConstrainingStructure}, the CMB temperature anisotropies are mainly generated by the coupled baryon-photon fluid. This fluid is normally affected separately by the neutrinos and the dark matter components, because of their different behaviour: that is, neutrinos free stream and dark matter clusters slowly. The solution to the system can be separated into slow modes and fast modes \cite{Weinberg2008,Voruz2013}, where the baryon-photon fluid and neutrinos fall into the former category and DM into the latter. This means that, in the non-interacting case, only the neutrinos will have significant gravitational interactions with the baryon-photon fluid. However, when the DM couples strongly enough to neutrinos, it will also experience oscillations and will contribute to the fast modes. This effect will  significantly affect the CMB spectrum, leading to a gravitational boosting effect, which enhances the peaks. Additionally, if the neutrinos and DM are tightly coupled, their lower sound speed compared to the baryon-photon fluid will cause a drag effect, leading to a lightly smaller acoustic scale, causing the CMB peaks to be shifted slightly towards higher multipoles $l$.

\begin{table}[]
    \centering
    \begin{tabular}{|c|c|}
    \hline
        Parameter & Prior \\
    \hline
        $\omega_{b}$ & Flat, unbounded \\
    \hline
        $\omega_{DM}$ & Flat, unbounded  \\
    \hline
        $100*\theta_s$ & Flat, unbounded  \\
    \hline
        $\ln10^{10}A_s$ & Flat, unbounded  \\
    \hline
        $n_s$& Flat, unbounded  \\
    \hline
        $\tau_{reio}$ & Flat, $\tau_{reio}\geq 0.004$  \\
    \hline
        $u_\chi$ & Flat, $u_\chi\geq 0$ \\
    \hline
        $\sum m_\nu $ & Flat, $\sum m_\nu \geq 0.06$ eV\\
    \hline
    \end{tabular}
    \caption{MCMC parameters and priors.}
    \label{tab:MCMCparams}
\end{table}

We run our MCMC chains varying the six usual $\Lambda$CDM cosmological parameters, i.e. the baryon and cold dark matter physical densities $\omega_{b} \equiv \Omega_b h^2$ and $\omega_{DM} \equiv \Omega_{DM} h^2$, the angular size of the sound horizon at last-scattering $100*\theta_s$, the optical depth to reionization $\tau$ and the amplitude and tilt of the primordial scalar power spectrum (represented by $\ln10^{10}A_s$ and $n_s$ respectively) in addition to the interaction parameter $u_\chi$. We have also run cases varying the sum of neutrino masses, assuming a degenerate mass hierarchy. While we know from neutrino oscillations that this is not the correct model, the difference will be negligible for our purposes \cite{Archidiacono2020WhatCosmology}. We use combinations of three different datasets for our MCMC. All our runs make use of the Planck 2018 CMB temperature plus high and low multipole polarization measurements, i.e the TTTEEE dataset (using the likelihoods highl\_TTTEEE, lowl\_TT, and lowl\_EE), which we refer to as \emph{Planck} or \emph{TTTEEE}. Some runs also include the Planck 2018 CMB lensing power spectrum reconstructed from the CMB temperature four-point function , referred to as \emph{Lensing} ~\cite{Aghanim:2018oex}. Finally, in some runs we shall also use BAO observations, specifically from BOSS DR12 \cite{Alam:2016hwk} and combined quasar and Lyman-$\alpha$ data from eBOSS DR14 \cite{Blomqvist2019,Cuceu_2019,Agathe:2019vsu}, denoted as \emph{BAO}.

Tables \ref{tab:plancktest} and \ref{tab:Mtest} show the best fit values with 95\%~CL limits for constant and varying neutrino mass, respectively. For $u_\chi$ and $\sum m_\nu$, 95\%~CL upper limits are shown. 
\input{table_plancktest}
\input{table_Mtest}

\begin{figure}
	\centering
	\includegraphics[width=0.5\linewidth]{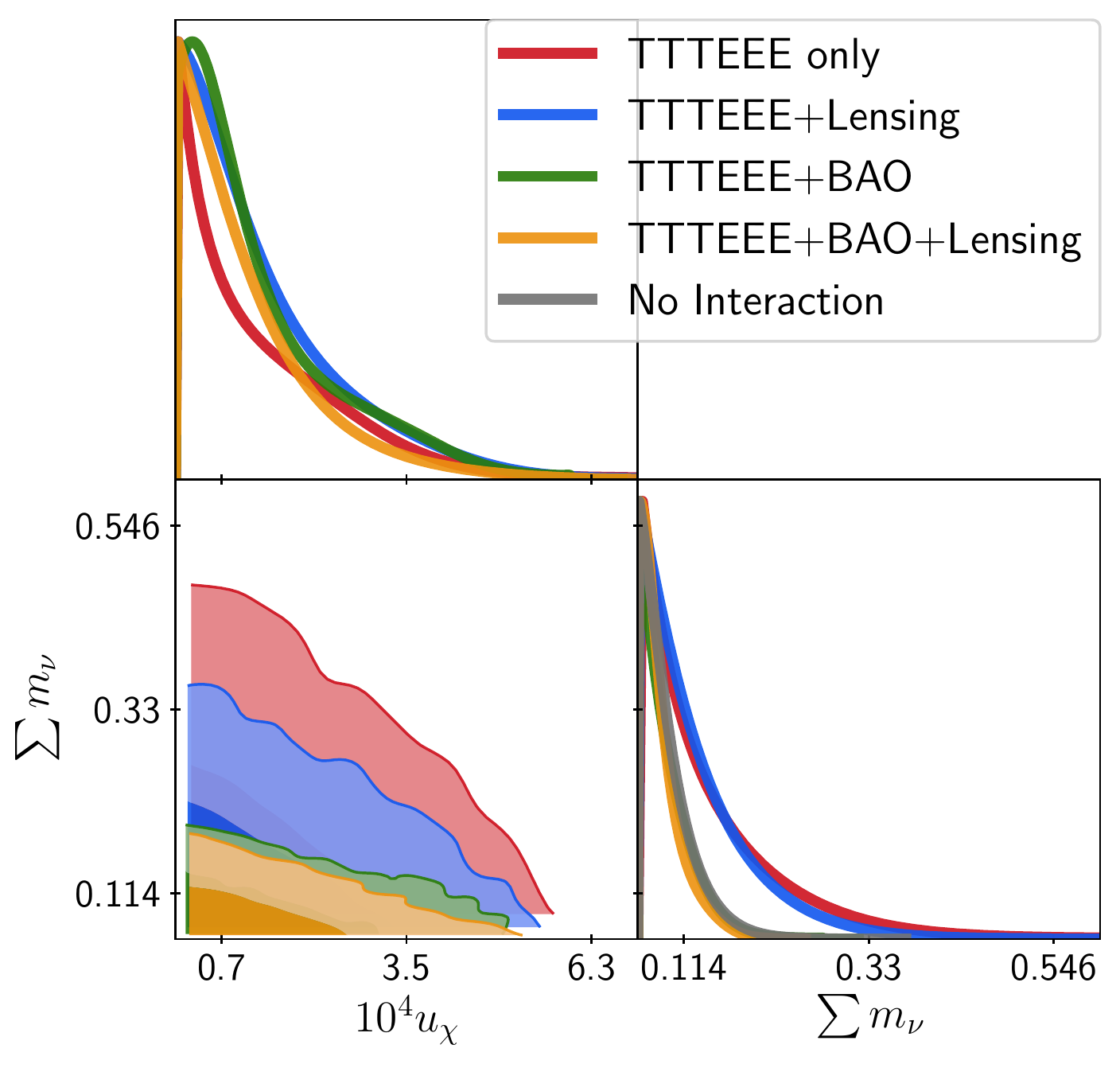}
	\caption{One-dimensional posterior probability distributions for $u_\chi$ and $\sum m_\nu$ for different combination of datasets and two-dimensional $68\%$ and $95\%$~CL allowed regions in the ($u_\chi$,$\sum m_\nu$) plane. The 'Non-Interacting' posterior uses all the three datasets, that is, Planck CMB TTTEEE+ Planck CMB Lensing + BAO.}
	\label{fig:u_mnu}
\end{figure}

We find the correlation between $u_\chi$ and most of the \LCDM parameters to be weak at most, but there seems to be a slight correlation with $\theta_s$ and anticorrelation with $n_s$. We show the posteriors for $u_\chi$ and $\sum m_\nu$ in figure \ref{fig:u_mnu}. The two-dimensional posterior has the characteristic quarter-ellipse shape, as data bounds both parameters from above. It is evident that there is a significant difference in the upper bound on the neutrino mass depending on whether BAO data is used, regardless of whether neutrinos are interacting or not. The interaction strength $u_\chi$ shows no such strong dependence on datasets. In Appendix \ref{sec:2dplots} we provide a list of figures illustrating all the correlations in two-dimensional plots, which also include the one-dimensional posterior probabilities.

Our 95\%~CL upper limits on the interaction cross section are slightly weaker than those previously found for massless neutrinos, see e.g.  \cite{DiValentino2018ReducingInteractions,Stadler2019How.,Diacoumis:2018ezi}. This was expected, as the main effect if the mass on the interaction term is through the $p^2/E^2$-term, which can only make the interaction weaker, allowing for a larger value of $u_\chi$. It is also worth noting that we have used a flat prior for $u_\chi$, while previous studies have sometimes used Jeffreys or logarithmic priors, which can make a difference\cite{Diacoumis:2018ezi}. An initial guess for the upper limit of $u_\chi$ was found by running a shorter chain with a logarithmic prior to obtain an appropriate scale for the MCMC step size and the starting point.

\begin{figure}
	\centering
	\includegraphics[width=0.5\linewidth]{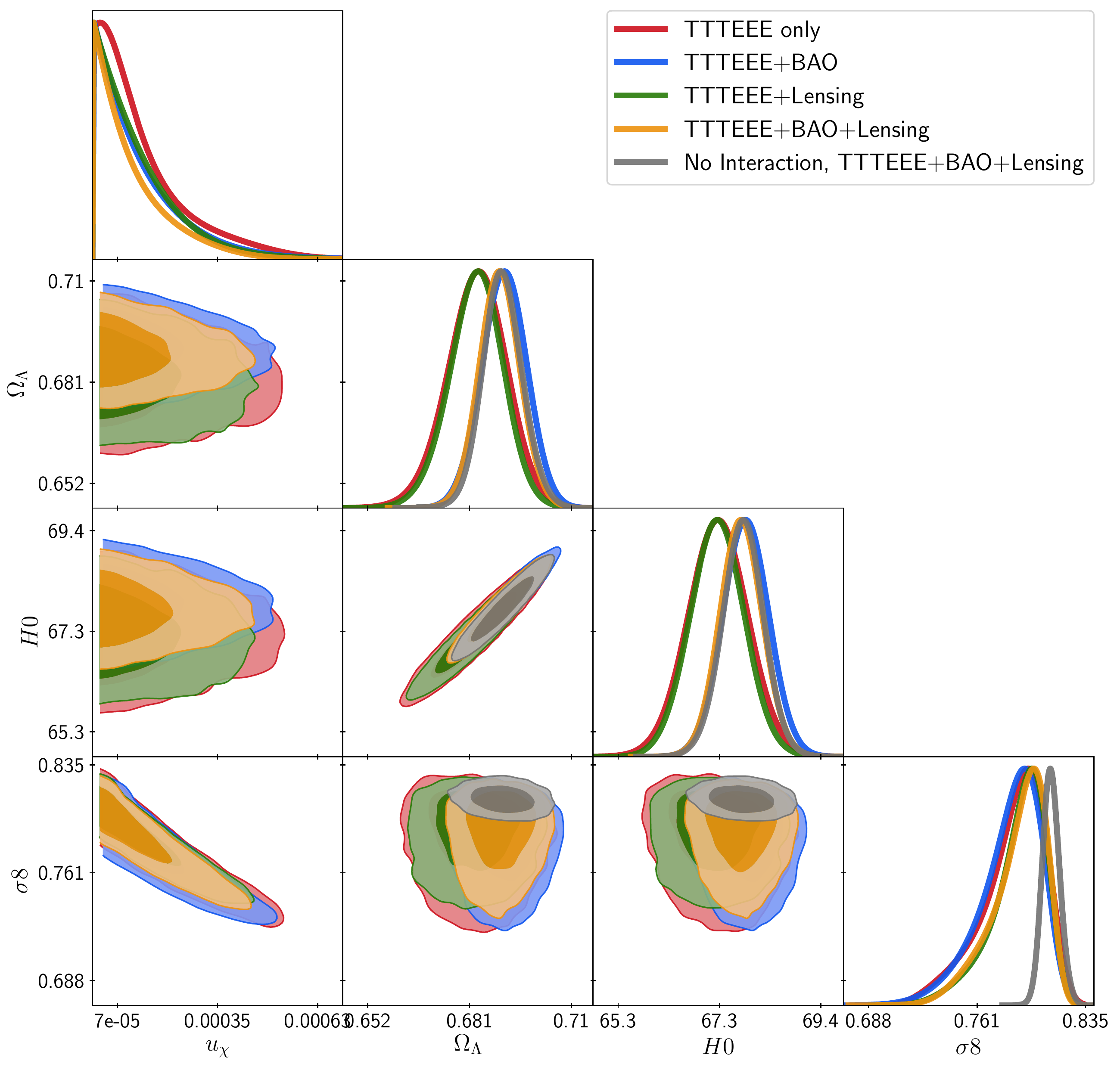}
	\caption{One-dimensional posterior probability distributions for the most interesting parameters for different combination of datasets and two-dimensional $68\%$ and $95\%$~CL allowed regions. Note the very strong correlation between $u_\chi$ and $\sigma_8$. In the interacting case, the viable range for $\sigma_8$ is significantly larger than in standard $\Lambda$CDM, see text for details.}
	\label{fig:H0_sigma8}
\end{figure}

Another interesting issue to explore is how the interaction affects $H_0$ and $\sigma_8$, as there is currently tension in the field concerning the values of these two parameters. They are calculated as derived parameters in our MCMC chains and their posterior distributions are depicted in figure~\ref{fig:H0_sigma8}. Note that there is no real correlation between $u_\chi$ and $H_0$, and therefore this interaction does not offer a solution to the $H_0$-tension. The main effect on $H_0$ comes from $\Omega_\Lambda$, which is set indirectly by enforcing flatness. Thus the lower  dark matter abundance in the chains using BAO data is compensated with an increase in $\Omega_\Lambda$ and by extension $H_0$. However, these differences are almost negligible. On the other hand, $\sigma_8$ shows a strong dependence on $u_\chi$, a stronger interaction yielding a lower $\sigma_8$, which may help in this tension, especially because this easing of the $\sigma_8$-tension comes without exacerbating the $H_0$-tension \cite{valentino2020sigma8}. 
The weak-lensing-devoted KiDS-1000 survey finds $\sigma_8 = 0.76^{+0.025}_{-0.020}$ \cite{KiDS-1000}, which is slightly lower than our best fit, but consistent within our $1\sigma$ lower limit for some of our data combinations and consistent with our $2\sigma$ lower limits for all data combinations, see tables~\ref{tab:plancktest} and \ref{tab:Mtest}. Comparing to the non-interacting case, it is evident that there is no change in $H_0$ in the allowed parameter space, but the allowed range  for $\sigma_8$ is significantly enlarged. 

\section{Conclusions}
\label{sec:conclusion}
We have presented a complete Boltzmann hierarchy for massive neutrinos interacting with cold dark matter, using a momentum independent scattering cross section (similar to e.g.\ Thomson scattering). As a novelty, we have also accounted for non-zero neutrino masses in the derivation, which means updating the Boltzmann hierarchy. We have implemented this interaction in the Boltzmann code CLASS, which we have used together with the MCMC sampler MontePython to infer upper limits for the interaction strength $u_\chi$, using Planck 2018 data \cite{Planck2018I,Planck2018} as well as SDSS BAO data from galaxies, Ly$\alpha$ and quasars~\cite{Blomqvist2019,Cuceu_2019,Agathe:2019vsu,Alam:2016hwk}. Our implementation yields results similar to previous work for the effect on the CMB and matter power spectra. However, we generally find slightly looser upper bounds for the interaction strength $u_\chi$ than previous work developed under the approximation of massless neutrinos. This is in good agreement with our expectations, since a new factor proportional to $p^2/E^2$ will appear in the interaction rate. This in turn suppresses the interaction compared to the case of massless neutrinos with the same $u_\chi$. A slightly larger $u_\chi$ is therefore needed to obtain the same effect on observables, yielding a slightly looser upper limit. 

We also find that the interaction does not appreciably change the upper bound on the sum of neutrino masses compared to the non-interacting case, meaning that the usual cosmological neutrino bounds are robust even in the presence of this type of non-standard interactions. Concerning the current cosmological tensions, we do not find any significant change to the value of the Hubble constant $H_0$ in the interacting case. However, very interestingly, the interaction case shows an appreciable decrease in the value of $\sigma_8$, bringing it more in line with the values reported by the weak lensing surveys KiDS-1000 \cite{KiDS-1000} and DES \cite{DES-1-galaxy-wl,DES-1-shear} among others, which report a significantly lower value than Planck. Here the interaction offers a hint to a potential solution, since it eases the $\sigma_8$-tension without exacerbating the $H_0$-tension, which is often an issue \cite{valentino2020sigma8}. Future cosmological observations will further constrain the neutrino and dark matter sector interactions, shedding light on the  microphysics of these two invisible components of our universe. 

\acknowledgments
The authors acknowledge the Sydney Informatics Hub and the use of the University of Sydney’s high performance computing cluster, Artemis, some numerical results presented in this work were obtained at the Centre for Scientific Computing, Aarhus.\footnote{\url{http://phys.au.dk/forskning/cscaa/}}. O.M. is supported by the Spanish grants FPA2017-85985-P, PROMETEO/2019/083 and by the European ITN project HIDDeN (H2020-MSCA-ITN-2019//860881-HIDDeN).

\clearpage
\appendix
\input{derivation}

\section{Fluid Approximation} 
\label{sec:fluidapprox}

The publicly available Bolztmann solver code CLASS uses a fluid approximation for the non cold dark matter-species \cite{Lesgourgues2011TheRelics}, which is used for massive neutrinos, to minimize the computational time. In that approximation, the full integral over momentum-space is not performed, instead the hierarchy is truncated at $l_\text{max}=2$ and the continuity, Euler, and shear equations are calculated directly instead of performing an integral over the momentum. Since the interaction rate $C_\chi$ is momentum-dependent, see Eq.~(\ref{equation:chi}), the expression must be adapted to use some approximation suited to this particular regime.

In the regime where CLASS uses the fluid approximation to evolve $\delta_\nu$, $\theta_\nu$, and $\sigma_\nu$, rather than the full momentum hierarchies, the interaction term acts both as a collision and a damping term, similarly to the Thomson scattering case. The shapes of these terms are independent of the gauge, yielding the modified equations:
\begin{equation*}
    \dot{\theta}_\nu = \left[...\right] + \widetilde{C}_\chi (w) \left( \theta_\chi - \theta_\nu\right), \qquad \dot{\sigma}_\nu = \left[...\right] - \frac{9}{10} \widetilde{C}_\chi (w) \sigma_\nu,
\end{equation*}
where $\left[...\right]$ represents the usual terms present in the non-interacting case, as described in \cite{Lesgourgues2011TheRelics}. The interaction rate does not change the number of neutrinos or facilitate any appreciable energy transport between species, so no modification to the $\delta_\nu$-equation is made.

Since the momentum dependence of the interaction rate is  $\propto \frac{p^2}{E^2}$, it is reasonable to assume that the integrated rate $\widetilde{C}_\chi$ is related to the equation of state parameter $w$, as both are integrated quantities dependent on the ratio between the  kinetic and total energies. However, the relationship between  $\widetilde{C}_\chi$ and $w$  is highly non-trivial. We extract it numerically by calculating both $\widetilde{C}_\chi$ and $w$ for a range of temperatures. On the other hand, such a relation turns out to be independent of the particle mass, mildly simplifying the interacting scenario. The relationship between  $\widetilde{C}_\chi$ and $w$ is well approximated by
\begin{equation}
    \widetilde{C}_\chi (w)= S_\chi \left[ A\,(3w)^B+Cw \right]~,
\end{equation}
where $S_\chi$ contains all the factors that do not depend on the momentum $p$, 
\begin{equation}
    S_\chi = a u_{\nu\chi} \frac{\sigma_{\text{Th}} \rho_\chi}{100\, \text{GeV}}.
\end{equation}
The parameters $A$, $B$, and $C$ have the following values:
\begin{equation*}
    A = 1.52116~, \quad B = 0.50884~, \quad C = -1.56422~,
\end{equation*}
which provides a very good fit. 

\begin{figure}
	\centering
	\includegraphics[width=0.8\linewidth]{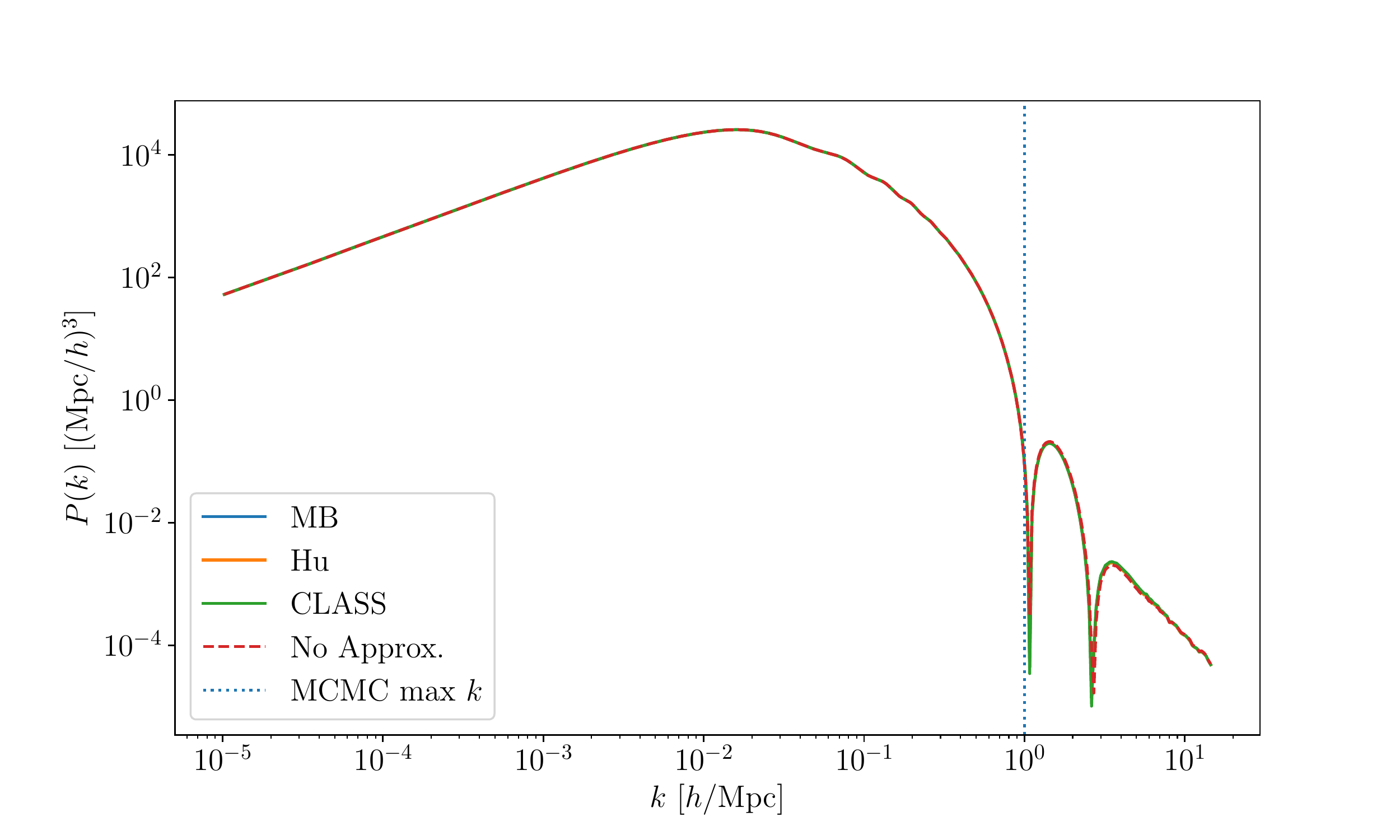}
	\caption{The matter power spectrum computed with the different prescriptions in in CLASS for evolving the shear, $\sigma_\nu$, in the truncated spectrum, as well as the case of using the full hierarchy. MB is the standard Ma and Bertschinger approach \cite{MaBertschinger}, Hu is an approach by Wayne Hu \cite{Hu:1998kj} and CLASS is a new approach introduced in the CLASS code, all three are detailed in references \cite{CLASSII,Lesgourgues2011TheRelics}. Cosmological parameters are set to their best fit value from MCMC with Planck 2018TTTEEE, except for $u_\chi=4.2\cdot10^{-4}$, which is set to its $2\sigma$ upper limit. It is clear that the approximations yield very similar results.}
	\label{fig:Pk_fluidcomp}
\end{figure}

The approximation introduces an error into the CMB and matter power spectra, which is important to quantize. The effect on the CMB power spectra is on the order of $0.01\%$, and therefore completely negligible. Concerning the matter power spectrum in the interacting scenario there is some error at very small scales. Nevertheless these scales are well beyond the linear regime probed here. The error on the matter power spectrum is mostly related to the exact shape of the oscillatory features, as illustrated in figure~\ref{fig:Pk_fluidcomp}. Note that the regime where the discrepancy is largest falls well outside the area probed by our MCMC, as we use a maximum scale $k_{max}=1$. The largest discrepancy is therefore $\sim 5\%$ at $k\rightarrow1$. In order to ensure that the error from the fluid approximation does not introduce a significant bias in our MCMC results, we have recalculated a few selected points using high-precision settings, without the fluid approximation, to probe that the $\chi^2$ shows the same behavior as the one obtained using the approximations, obtaining insignificant changes to the likelihood. We therefore conclude that the error with normal precision settings remains within very acceptable limits.

\section{Parameter correlations}
\label{sec:2dplots}
\begin{figure}
	\centering
	\makebox[\textwidth][c]{\includegraphics[width=1.3\linewidth]{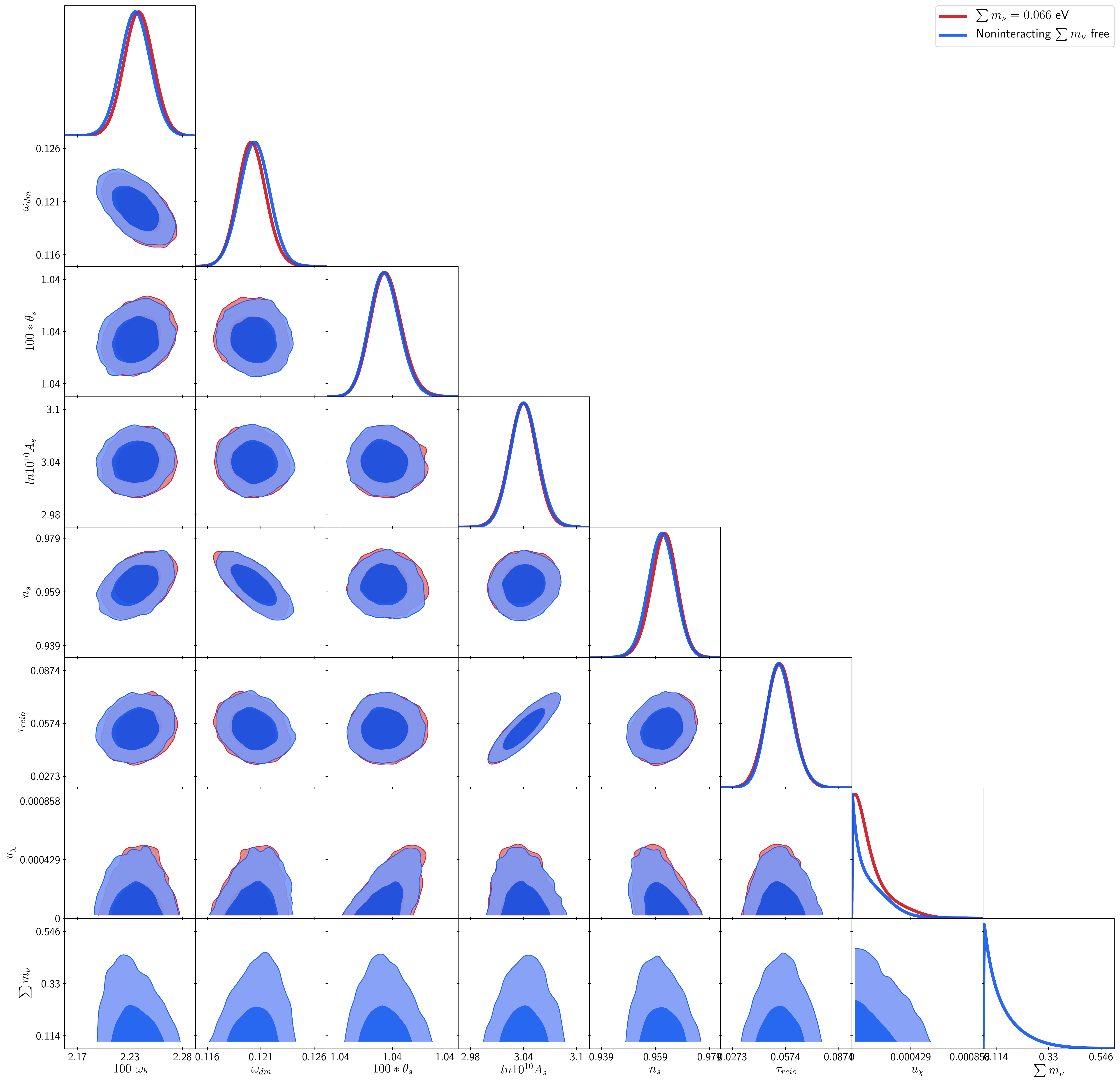}}
	\caption{One and two-dimensional probability posterior distributions computed using exclusively Planck TTTEEE data, comparing the results for a fixed neutrino mass $\sum m_\nu = 0.066$~eV and a freely varying neutrino mass with the prior $\sum m_\nu \geq 0.06$ eV.}
	\label{fig:m066vsmtest}
\end{figure}

\begin{figure}
	\centering
	\makebox[\textwidth][c]{\includegraphics[width=1.3\linewidth]{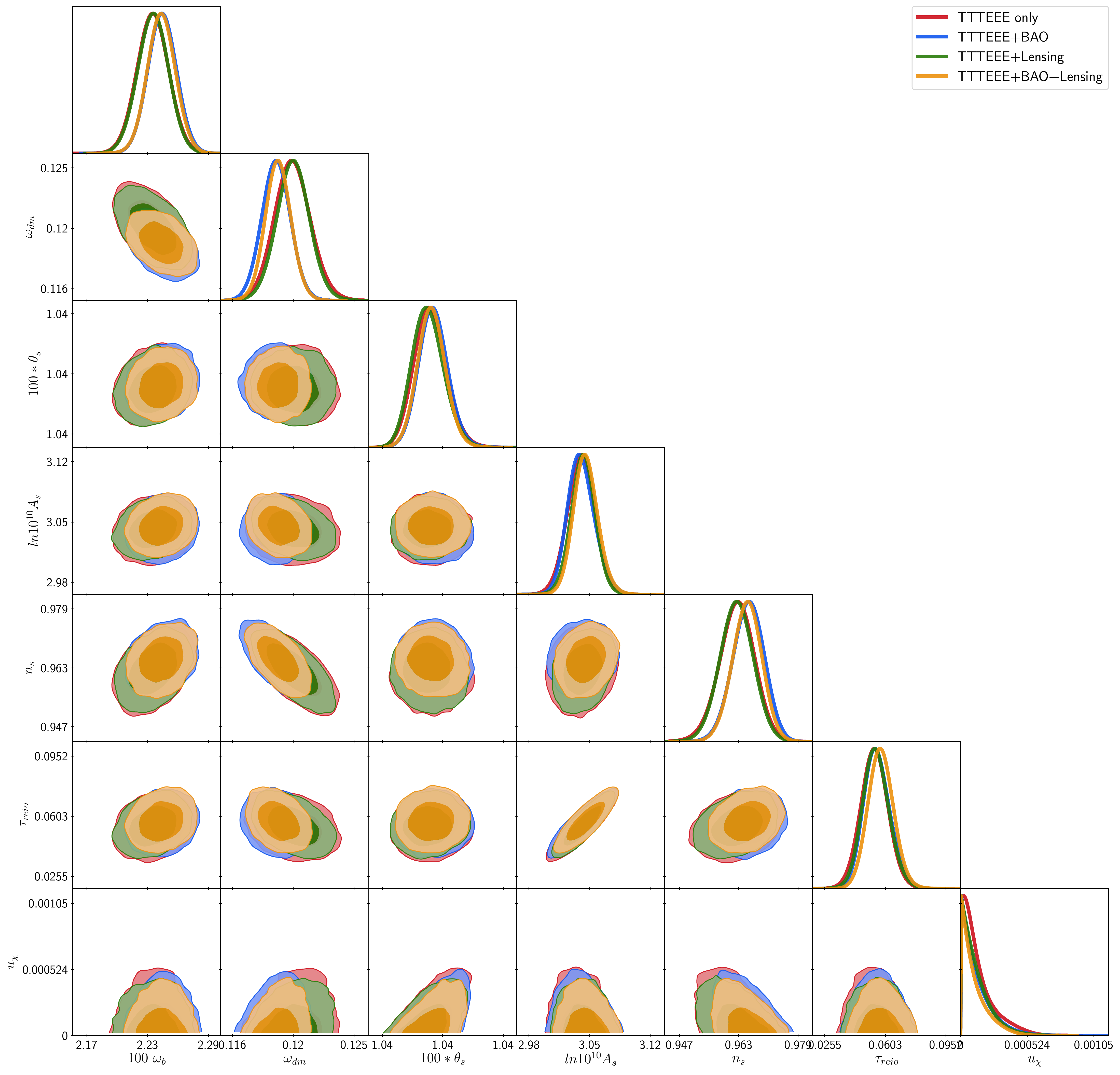}}
	\caption{One and two-dimensional probability posterior distributions computed using  a fixed neutrino mass $\sum m_\nu = 0.066$~eV for different combinations of datasets.}
	\label{fig:lowmass_alldata}
\end{figure}

\begin{figure}
	\centering
	\makebox[\textwidth][c]{\includegraphics[width=1.3\linewidth]{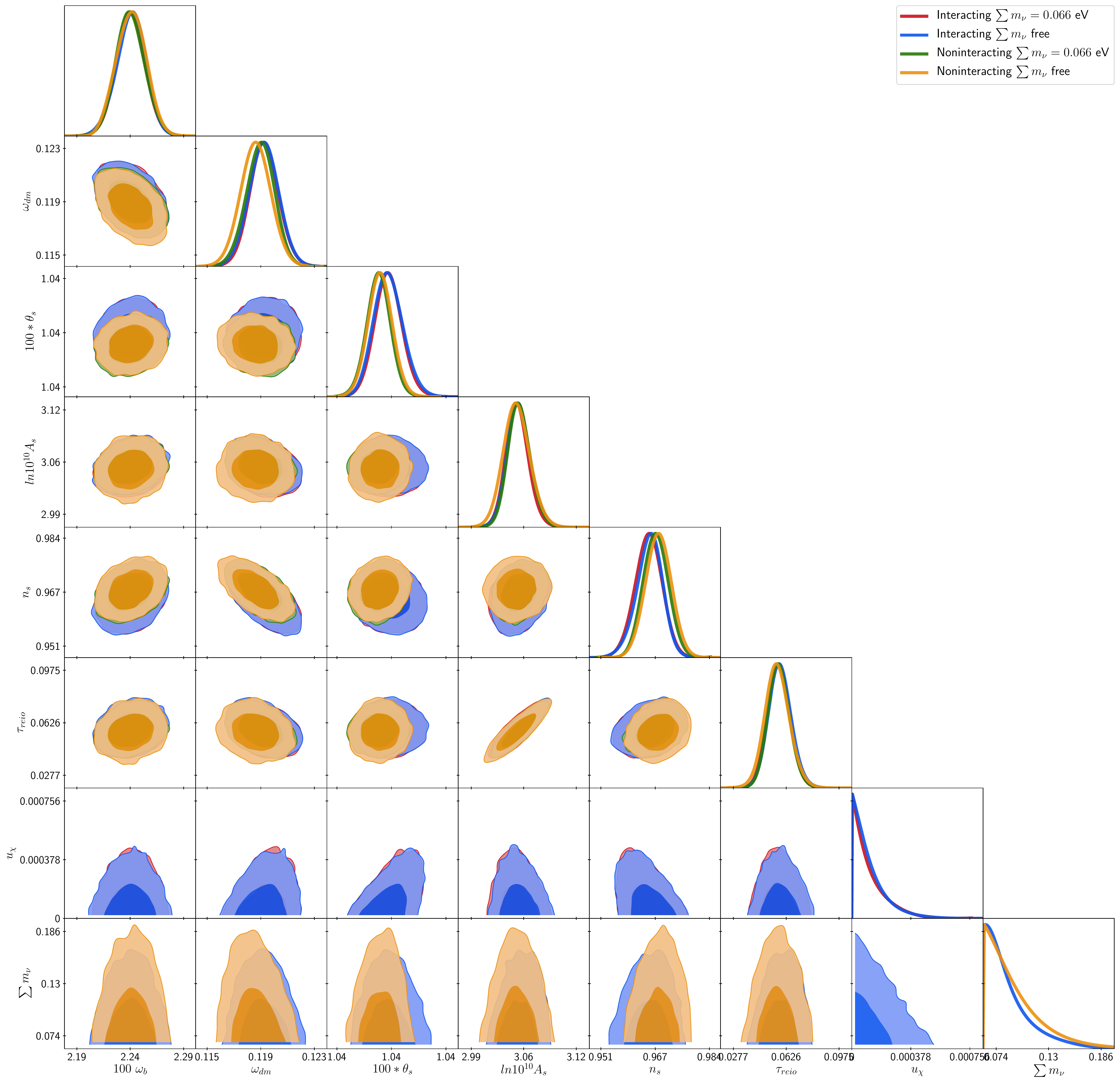}}
	\caption{Comparison of the one and two-dimensional probability posterior distributions using  a fixed neutrino mass $\sum m_\nu = 0.066$~eV and using $\sum m_\nu$ as a free parameter, for the interacting and non-interacting scenarios. These are computed using our full range of datasets, \emph{TTTEEE}+\emph{BAO}+\emph{lensing}. }
	\label{fig:mfree}
\end{figure}

For completeness, we present here the two-dimensional contour plots and the one-dimensional posterior distribution plots from our MCMC runs showing all varied parameters, see figures~\ref{fig:m066vsmtest}, \ref{fig:lowmass_alldata}, and \ref{fig:mfree}.
Figure \ref{fig:m066vsmtest} shows the posterior distributions computed using Planck TTTEEE data exclusively, comparing the results obtained after fixing the neutrino to $\sum m_\nu = 0.066$~eV to the results assuming  a free varying neutrino mass with the prior $\sum m_\nu \geq 0.06$~eV. The latter yields a $2\sigma$ upper limit of $\sum m_\nu \leq 0.339$~eV. A clear correlation between the interaction strength $u_\chi$ and several cosmological parameters, most notably $\theta_s$ and $n_s$, can be noticed from this figure .

Figure~\ref{fig:lowmass_alldata} shows the case of a fixed neutrino mass $\sum m_\nu = 0.066$~eV, using different combinations of datasets to constrain parameters. The general trends are similar to the non-interacting case, the addition of BAO data lowers the estimated value of $\omega_{DM}$, while raising $\omega_b$ and $n_s$, however, these shifts are not significant. It is interesting to note that the addition of lensing data helps in constraining $u_\chi$ more strongly. Nevertheless the constraints on the cosmological parameters are generally very similar, which may also indicate that they are dominated by their determination from Planck 2018 TTTEEE data.

Finally, figure~\ref{fig:mfree} compares the cases of interacting, non-interacting, fixed $m_\nu$ and free $m_\nu$. Note that the interacting case mildly favors a larger value of $\theta_s$ and a lower value of $n_s$. This also matches what can be expected from their correlations with $u_\chi$. Results for the non-interacting case were obtained by running chains with the same parameters and datasets, but setting $u_\chi$ to 0.

\clearpage

\bibliography{references}

\end{document}

%% file: derivation_commands.tex
\newcommand{\intdnu}[1]{\frac{d^3 #1}{\left(2\pi\right)^3 E_\nu\!\left(#1\right)}}

\newcommand{\intdnutwo}[1]{\frac{d^3 #1}{\left(2\pi\right)^3 2E_\nu\!\left(#1\right)}}

\newcommand{\intdc}[1]{\frac{d^3 #1}{\left(2\pi\right)^3 E_\chi\!\left(#1\right)}}

\newcommand{\intdctwo}[1]{\frac{d^3 #1}{\left(2\pi\right)^3 2E_\chi\!\left(#1\right)}}

\newcommand{\gq}[1]{g\!\left(\mathbf{q}#1\right)}

\newcommand{\fp}[1]{f\!\left(\mathbf{p}#1\right)}

\newcommand{\Enup}[1]{E_\nu\!\left(\mathbf{p}#1\right)}
\newcommand{\Echi}[1]{E_\chi\!\left(#1\right)}

\newcommand{\Ep}{E_\nu\!\left(p\right)}

\newcommand{\qpr}{\mathbf{q}+\mathbf{p}-\mathbf{p}'}
\newcommand{\gqpr}{g\!\left(\mathbf{q}+\mathbf{p}-\mathbf{p}'\right)}

\newcommand{\qb}{\mathbf{q}}
\newcommand{\pb}{\mathbf{p}}
\newcommand{\mv}{m_\chi\mathbf{v}_\chi}

\newcommand{\fz}[1]{f^{(0)}\!\left(p#1\right)}
\newcommand{\fo}[1]{f^{(1)}\!\left(\mathbf{p}#1\right)}

%% file: table_plancktest.tex
\begin{table}
\centering
\makebox[\textwidth][c]{
\begin{tabular}{|l|llll|}
\hline
                    & Planck TTTEEE                       & Planck + Lensing                     & Planck + BAO                         & \specialcell{Planck +\\ Lensing + BAO} \\ 
\hline
$100 \, \omega_b$   & $2.25^{+0.02}_{-0.04}$     & $2.23^{+0.03}_{-0.03}$     & $2.23^{+0.04}_{-0.01}$     & $2.25^{+0.02}_{-0.03}$                                      \\ 
\hline
$\omega_{DM}$       & $0.120^{+0.003}_{-0.002}$    & $0.121^{+0.002}_{-0.003}$    & $0.119^{+0.00}_{-0.002}$   & $0.119^{+0.002}_{-0.002}$                                 \\ 
\hline
$100 \, \theta_s$   & $1.0416^{+0.0014}_{-0.0001}$  & $1.0421^{+0.0008}_{-0.0007}$  & $1.0418^{+0.0012}_{-0.0002}$  & $1.0419^{+0.0010}_{-0.0003}$                                \\ 
\hline
$\ln 10^{10}A_{s}$  & $3.05^{+0.03}_{-0.04}$     & $3.04^{+0.03}_{-0.03}$      & $3.04^{+0.04}_{-0.03}$       & $3.04^{+0.04}_{-0.02}$                                    \\ 
\hline
$n_{s } $           & $0.963^{+0.009}_{-0.011}$    & $0.964^{+0.008}_{-0.011}$   & $0.970^{+0.004}_{-0.014}$   & $0.966^{+0.007}_{-0.010}$                                 \\ 
\hline
$\tau_{reio }$      & $0.056^{+0.015}_{-0.018}$   & $0.051^{+0.020}_{-0.012}$    & $0.055^{+0.017}_{-0.014}$    & $0.056^{+0.018}_{-0.013}$                                  \\ 
\hline
$u_{\chi} $         & $4.21\cdot10^{-4}$                & $3.55\cdot10^{-4}$                 & $3.85\cdot10^{-4}$                 & $3.40\cdot10^{-4}$                                               \\ 
\hline
$H_0$ [km/s/Mpc]    & $67.4^{+1.1}_{-1.4}$          & $67.1^{+1.3}_{-0.9}$          & $67.8^{+1.0}_{-0.9}$      & $67.8^{+0.8}_{-1.0}$                                     \\ 
\hline
$\sigma_8$          & $0.81^{+0.01}_{-0.07}$    & $0.80^{+0.02}_{-0.05}$     & $0.80^{+0.02}_{-0.06}$     & $0.80^{+0.02}_{-0.06}$          \\              \hline           
\end{tabular}
}
\caption{Best fit values with 95\% confidence limits for the case of constant neutrino mass, except for $u_\chi$, where only the 95\%~CL upper limit is shown.}
\label{tab:plancktest}
\end{table}

%% file: table_Mtest.tex
\begin{table}
\centering
\makebox[\textwidth][c]{
\begin{tabular}{|l|llll|}
\hline
                                    & Planck TTTEEE                       & Planck + Lensing                     & Planck + BAO                         & \specialcell{Planck +\\ Lensing + BAO}                         \\ 
\hline
$100 \, \omega_b$                   & $2.24^{+0.03}_{-0.04}$     & $2.24^{+0.03}_{-0.03}$     & $2.25^{+0.03}_{-0.03}$      & $2.24^{+0.03}_{-0.03}$       \\ 
\hline
$\omega_{DM}$                       & $0.120^{+0.003}_{-0.003}$ & $0.120^{+0.004}_{-0.001}$   & $0.120^{+0.002}_{-0.003}$   &$0.119^{+0.002}_{-0.002}$   \\ 
\hline
$100 \, \theta_s$                   & $1.0420^{+0.0009}_{-0.0005}$ & $1.0419^{+0.0010}_{-0.0005}$ & $1.0419^{+0.0011}_{-0.0004}$ & $1.0419^{+0.0010}_{-0.0004}$ \\ 
\hline
$\ln 10^{10}A_{s}$                  & $3.05^{+0.03}_{-0.04}$    & $3.04^{+0.04}_{-0.02}$      & $3.03^{+0.05}_{-0.02}$      & $3.05^{+0.03}_{-0.03}$     \\ 
\hline
$n_{s } $                           & $0.963^{+0.009}_{-0.012}$  & $0.965^{+0.006}_{-0.014}$  & $0.966^{+0.008}_{-0.009}$   & $0.967^{+0.007}_{-0.010}$   \\ 
\hline
$\tau_{reio }$                      & $0.055^{+0.016}_{-0.016}$  & $0.0528^{+0.019}_{-0.012}$  & $0.048^{+0.026}_{-0.006}$   & $0.057^{+0.017}_{-0.014}$    \\ 
\hline
$u_{\chi} $                         & $3.97\cdot10^{-4}$                & $3.83\cdot10^{-4}$                 & $3.83\cdot10^{-4}$                 & $3.34\cdot10^{-4}$        \\ 
\hline
$\sum m_\nu$ [eV]                   & $0.33$                           & $0.26$                            & $0.15$                            & $0.14$                                     \\ 
\hline
$H_0$ [km/s/Mpc]                    & $67.2^{+1.2}_{-3.3}$        & $67.3^{+0.9}_{-2.9}$          & $67.5^{+1.2}_{-0.9}$         & $67.6^{+1.0}_{-1.0}$      \\ 
\hline
$\sigma_8$                          & $0.80^{+0.01}_{-0.09}$   & $0.79^{+0.03}_{-0.06}$    & $0.80^{+0.02}_{-0.07}$    & $0.81^{+0.01}_{-0.06}$   \\
\hline
\end{tabular}
}
\caption{Best fit values with 95\% confidence limits for the case of varying neutrino mass, except for $u_\chi$ and $\sum m_\nu$, where 95\%~CL upper limits are shown.}
\label{tab:Mtest}
\end{table}

%% file: derivation.tex
\section{Derivation}
\label{sec:derivation}
We assume that the DM-particle is much heavier than the neutrino and completely nonrelativistic. This means that we can use approximations similar to those used when deriving the Thompson scattering terms.

The neutrino distribution function obeys the Boltzmann equation
\begin{equation}
	\frac{d}{dt} f \left(\mathbf{x},\mathbf{p},t\right) = C\left[f \left(\mathbf{x},\mathbf{p},t\right)\right].
\end{equation}
The left-hand side has been well described in the literature. In the standard case of noninteracting massive neutrinos, the right-hand side is zero, but when introducing an interaction with dark matter, a collision term appears. The form of this term will be derived in the following section. We will proceed in a fashion very similar to the derivation of the first-order Thompson scattering term in \cite{Dodelson1995ReionizationBackground}.

\subsection{The Collision Term}
We do not assume any quantum statistics for the DM particles, and thus do not introduce any Pauli-blocking or stimulated emission terms for them. The collision term is then determined by the integral
\begin{multline}
	C(p) = \frac{1}{E_\nu \! \left(\mathbf{p}\right)}\int \intdnutwo{\mathbf{p}'} \intdctwo{\mathbf{q}} \intdctwo{\mathbf{q}'} \left(2\pi\right)^4 \left|M\right|^2 \\
	\times\delta^4 \! \left(q + p - q' - p'\right) \left[\gq{'}\fp{'}\left(1-\fp{}\right)-\gq{}\fp{}\left(1-\fp{'}\right)\right],
\end{multline}
where $p$ and $q$ denote the momenta of neutrinos and DM particles respectively. Unprimed quantities are initial state values and primed quantities are final state values. $f$ and $g$ are the distribution functions of neutrinos and DM particles respectively and $E_i$ is the total energy of a particle of species $i$, given as usual by
\begin{equation}
	E_i = \sqrt{m_i^2 + p_i^2}~.
\end{equation}
Finally $\left|M\right|^2$ is the matrix element squared of the interaction.

We perform the $\mathbf{q}'$ integral using the spatial part of the $\delta$-function, yielding
\begin{multline}
C(p) = \frac{1}{8E_\nu \! \left(\mathbf{p}\right)}\int \intdnu{\mathbf{p}'} \intdc{\mathbf{q}} \\ \times\frac{\left(2\pi\right)^4\left|M\right|^2}{\left(2\pi\right)^3\Echi{\qpr}} 
\delta \! \left(\Echi{\qb} + \Enup{} - \Echi{\qpr} - \Enup{'}\right)\\ 
\times\left[\gqpr\fp{'}\left(1-\fp{}\right)-\gq{}\fp{}\left(1-\fp{'}\right)\right], 
\end{multline}
Because the DM particles are so much heavier than the neutrinos, the energy transfer will be very small. We expand the DM energies in the $\delta$-function. To the first order,
\begin{equation}
	\Echi{\qb} = \sqrt{m_\chi^2 + q^2} \simeq m_\chi + \frac{q^2}{2m_\chi},
\end{equation}
thus
\begin{multline}
	\delta \! \left(\Echi{\qb} + \Enup{} - \Echi{\qpr} - \Enup{'}\right) \\
	\simeq 	\delta \! \left(m_\chi + \frac{q^2}{m_\chi} + \Enup{} - m_\chi - \frac{q^2}{m_\chi} - \frac{\left(\pb -\pb'\right)^2}{2m_\chi} -\frac{\left(\pb -\pb'\right)\cdot\qb}{m_\chi} - \Enup{'}\right)\\
	=\delta \! \left(\Enup{}- \Enup{'} - \frac{\left(\pb -\pb'\right)^2}{2m_\chi} -\frac{\left(\pb -\pb'\right)\cdot\qb}{m_\chi} \right).
\end{multline}
We then expand the $\delta$-function in terms of the energy transfer. To the first order,
\begin{equation}
	\begin{split}
	\delta \! \left(\Enup{}- \Enup{'} - \frac{\left(\pb -\pb'\right)^2}{2m_\chi} -\frac{\left(\pb -\pb'\right)\cdot\qb}{m_\chi} \right)
	=&\delta \! \left(\Enup{}- \Enup{'}\right)\\
	&+\left[ \frac{\left(\pb -\pb'\right)^2}{2m_\chi} +\frac{\left(\pb -\pb'\right)\cdot\qb}{m_\chi}\right]\\
	&\times \frac{\partial\delta \! \left(\Enup{}- \Enup{'}\right)}{\partial\Enup{'}}.
	\end{split}
\end{equation}
We can then change the variable of differentiation to $p'$, using
\begin{equation*}
	\frac{d}{dp} \Enup{} = \frac{p}{\Enup{}},
\end{equation*}
so
\begin{equation}
	\frac{\partial\delta \! \left(\Enup{}- \Enup{'}\right)}{\partial\Enup{'}} = \frac{\partial\delta \! \left(\Enup{}- \Enup{'}\right)}{\partial p'} \frac{\Enup{'}}{p'}.
\end{equation}
The collision integral is then expressed as
\begin{multline}
C(p) = \frac{1}{8E_\nu \! \left(\mathbf{p}\right)}\int \intdnu{\mathbf{p}'} \intdc{\mathbf{q}} \frac{\left(2\pi\right)^4\left|M\right|^2}{\left(2\pi\right)^3\Echi{\qpr}}\\
\times\left\{\delta \! \left(\Enup{} - \Enup{'}\right)+ \left[ \frac{\left(\pb -\pb'\right)^2}{2m_\chi} +\frac{\left(\pb -\pb'\right)\cdot\qb}{m_\chi}\right]\frac{\partial\delta \! \left(\Enup{}- \Enup{'}\right)}{\partial p'} \frac{\Enup{'}}{p'}\right\}\\ 
\times\left[\gqpr\fp{'}\left(1-\fp{}\right)-\gq{}\fp{}\left(1-\fp{'}\right)\right].
\end{multline}

Under the assumption that the DM temperature is extremely low, the distribution function is
\begin{equation}
	\gq{} \simeq n_\chi \left(2\pi\right)^3 \delta^3 \! \left(\qb - m_\chi\mathbf{v}_\chi\right),
\end{equation}
fulfilling the normalisation condition that
\begin{equation*}
	n_\chi = \int \frac{d^3 \qb}{\left(2\pi\right)^3} \gq{}.
\end{equation*}
This allows only two values of $\qb$ to give nonzero contributions to the collision integral:
\begin{align*}
	\gq{} &\rightarrow \qb = \mv,\\
	\gqpr &\rightarrow \qb = \mv + \pb' - \pb.
\end{align*}

The collision integral is then expressed as 
\begin{equation}
	C(p) = \frac{1}{8E_\nu \! \left(\mathbf{p}\right)}\int \intdnu{\mathbf{p}'} \frac{\left(2\pi\right)^4\left|M\right|^2}{\left(2\pi\right)^3\Echi{\mv}} \bigg[A+B\bigg]~,
\end{equation}
with the $A$ term corresponding to $\qb = \mv$ and the $B$ term corresponding to $\qb = \mv + \pb' - \pb$. Then
\begin{equation}
	\begin{split}
	A = &\frac{1}{\Echi{\mv + \pb - \pb'}}\\
	&\times\left\{\delta \! \left(\Enup{} - \Enup{'}\right)+ \left[ \frac{\left(\pb -\pb'\right)^2}{2m_\chi} +\frac{\left(\pb -\pb'\right)\cdot\mv}{m_\chi}\right]\frac{\partial\delta \! \left(\Enup{}- \Enup{'}\right)}{\partial p'} \frac{\Enup{'}}{p'}\right\}\\
	&\times\left[g\!\left(\mv+\pb-\pb'\right)\fp{'}\left(1-\fp{}\right)-n_\chi\fp{}\left(1-\fp{'}\right)\right]~,
	\end{split}
\end{equation}
and
\begin{equation}
	\begin{split}
	B = &\frac{1}{\Echi{\mv - \pb + \pb'}}\\
	&\times\left\{\delta \! \left(\Enup{} - \Enup{'}\right)+ \left[ \frac{\left(\pb -\pb'\right)^2}{2m_\chi} +\frac{\left(\pb -\pb'\right)\cdot\left(\mv+ \pb' - \pb\right)}{m_\chi}\right]\frac{\partial\delta \! \left(\Enup{}- \Enup{'}\right)}{\partial p'} \frac{\Enup{'}}{p'}\right\}\\
	&\times\left[n_\chi\fp{'}\left(1-\fp{}\right)-g\!\left(\mv-\pb+\pb'\right)\fp{}\left(1-\fp{'}\right)\right].
	\end{split}
\end{equation}
We throw away all terms $\propto \frac{1}{m_\chi}$, leading to
\begin{multline}
	C(p) = \frac{1}{8E_\nu \! \left(\mathbf{p}\right)}\int \intdnu{\mathbf{p}'} \frac{\left(2\pi\right)^4\left|M\right|^2}{\left(2\pi\right)^3\Echi{\mv}}\\
	\times\left\{\delta \! \left(\Enup{} - \Enup{'}\right)+ \left(\pb -\pb'\right)\cdot\mathbf{v}_\chi\frac{\Enup{'}}{p'}\frac{\partial\delta \! \left(\Enup{}- \Enup{'}\right)}{\partial p'} \right\}\\
	\times\bigg[\frac{1}{\Echi{\mv + \pb - \pb'}} \left[g\!\left(\mv+\pb-\pb'\right)\fp{'}\left(1-\fp{}\right)-n_\chi\fp{}\left(1-\fp{'}\right)\right]\\
	+\frac{1}{\Echi{\mv - \pb + \pb'}}\left[n_\chi\fp{'}\left(1-\fp{}\right)-g\!\left(\mv-\pb+\pb'\right)\fp{}\left(1-\fp{'}\right)\right]\bigg].
\end{multline}
We split the neutrino distribution function is a zeroth-order and a first-order part:
\begin{equation*}
	\fp{} = \fz{}+\fo{},
\end{equation*}
where the zeroth-order part is independent of the momentum direction. Hence, 
\begin{multline}
C(p) = \frac{1}{8E_\nu \! \left(\mathbf{p}\right)}\int \intdnu{\mathbf{p}'} \frac{\left(2\pi\right)^4\left|M\right|^2}{\left(2\pi\right)^3\Echi{\mv}}\\
\times\left\{\delta \! \left(\Enup{} - \Enup{'}\right)+ \left(\pb -\pb'\right)\cdot\mathbf{v}_\chi\frac{\Enup{'}}{p'}\frac{\partial\delta \! \left(\Enup{}- \Enup{'}\right)}{\partial p'} \right\}\\
\times\bigg[\frac{1}{\Echi{\mv + \pb - \pb'}} \left[g\!\left(\mv+\pb-\pb'\right)\left\{ \fz{'}+\fo{'} \right\}\left(1-\left\{ \fz{}+\fo{} \right\}\right)\right.\\
\left.-n_\chi\left\{ \fz{}+\fo{} \right\}\left(1-\left\{ \fz{'}+\fo{'} \right\}\right)\right]\\
+\frac{1}{\Echi{\mv - \pb + \pb'}}\left[n_\chi\left\{ \fz{'}+\fo{'} \right\}\left(1-\left\{ \fz{}+\fo{} \right\}\right)\right.\\
\left.-g\!\left(\mv-\pb+\pb'\right)\left\{ \fz{}+\fo{} \right\}\left(1-\left\{ \fz{'}+\fo{'} \right\}\right)\right]\bigg].
\end{multline}
For the sake of simplicity, this can be written in terms of for phase-space terms, $\delta1,\,\delta2,\,\partial\delta1,\,\partial\delta2$,
\begin{multline}
C(p) = \frac{1}{8E_\nu \! \left(\mathbf{p}\right)}\int \intdnu{\mathbf{p}'} \frac{\left(2\pi\right)^4\left|M\right|^2}{\left(2\pi\right)^3\Echi{\mv}}\bigg[\\
\delta \! \left(\Enup{} - \Enup{'}\right) \left[\delta1 + \delta2\right]\\
+ \left(\pb -\pb'\right)\cdot\mathbf{v}_\chi\frac{\Enup{'}}{p'}\frac{\partial\delta \! \left(\Enup{}- \Enup{'}\right)}{\partial p'} \left[\partial\delta1 + \partial\delta2\right] \bigg].
\end{multline}
The only terms multiplied with the $\delta$-function that will be nonzero, are those that fulfil the condition $\Enup{} = \Enup{'}$, hence $\left|\pb\right| = \left|\pb'\right|$ and $\fz{'}\rightarrow\fz{}$. Since, the $\delta$-function is a zeroth-order term, zeroth-order and first-order terms are kept in $\delta1$ and $\delta2$.

Writing out $\delta1$,
\begin{multline}
	\delta1 = \frac{1}{\Echi{\mv + \pb - \pb'}} \bigg[g\!\left(\mv + \pb - \pb'\right)\\\left(\fz{} + \fo{'} - \left(\fz{}\right)^2 - \fo{'}\fo{} - \fz{}\left(\fo{}+\fo{'}\right) \right)\\
	-n_\chi\left(\fz{}+\fo{} - \left(\fz{}\right)^2 - \fo{}\fo{'} - \fz{}\left(\fo{}+\fo{'}\right) \right)\bigg],
\end{multline}
the $\fo{}\fo{'}$-terms are discarded, since they are second-order. It is evident that if $\pb=\pb'$, the two terms are identical, leading to $\delta1=0$. For all other values of $\pb'$, we have $g\!\left(\mv + \pb - \pb'\right)=0$, thus the nonzero contributions yield
\begin{equation}
	\delta1 = \frac{n_\chi}{\Echi{\mv + \pb - \pb'}}\left(
	\left(\fz{}\right)^2 + \fz{}\left(\fo{}+\fo{'}\right)  - \fz{} - \fo{}
	\right).
\end{equation}
Using the same arguments, we get
\begin{equation}
	\delta2 = \frac{n_\chi}{\Echi{\mv - \pb + \pb'}}\left(
	\fz{} + \fo{'} -\left(\fz{}\right)^2 - \fz{}\left(\fo{}+\fo{'}\right)
	\right).
\end{equation}
The $\partial\delta1$ and $\partial\delta2$ terms are multiplied with a first-order term, hence, only their zeroth-order terms are kept. However, there is no $\delta$-function to set $\left|\pb\right| = \left|\pb'\right|$. By the same arguments as used for $\delta1$ and $\delta2$ we obtain
\begin{align}
	\partial\delta1 &= -\frac{n_\chi}{\Echi{\mv + \pb - \pb'}}\left(
	\fz{}-\fz{}\fz{'}
	\right),\\
	\partial\delta2 &= \frac{n_\chi}{\Echi{\mv - \pb + \pb'}}\left(
	\fz{'}-\fz{}\fz{'}.
	\right)
\end{align}

Since the DM particles are very nonrelativistic, we can make the approximation that $E_\chi\simeq m_\chi$. This allows us to reduce the terms:
\begin{align}
	\delta1+\delta2 &= \frac{n_\chi}{m_\chi}\left(\fo{'}-\fo{}\right)\\
	\partial\delta1+\partial\delta2 &= \frac{n_\chi}{m_\chi}\left(\fz{'}-\fz{}\right).
\end{align}

We can then write the collision integral as
\begin{multline}
	C(p) = \frac{n_\chi}{8E_\nu \! \left(\mathbf{p}\right)m_\chi^2}\int \intdnu{\mathbf{p}'} \frac{\left(2\pi\right)^4\left|M\right|^2}{\left(2\pi\right)^3}\bigg[\\
	\delta \! \left(\Enup{} - \Enup{'}\right) \left[\fo{'}-\fo{}\right]\\
	+ \left(\pb -\pb'\right)\cdot\mathbf{v}_\chi\frac{\Enup{'}}{p'}\frac{\partial\delta \! \left(\Enup{}- \Enup{'}\right)}{\partial p'} \left[\fz{'}-\fz{}\right] \bigg].
\end{multline}

We then split up the integral in an angular and a magnitude part
\begin{multline}
	C(p) = \frac{n_\chi}{8E_\nu \! \left(\mathbf{p}\right)m_\chi^2}\int \frac{dp' d\Omega p'^2}{\left(2\pi\right)^3 \Enup{'}} \frac{\left(2\pi\right)^4\left|M\right|^2}{\left(2\pi\right)^3}\bigg[\\
	\delta \! \left(\Enup{} - \Enup{'}\right) \left[\fo{'}-\fo{}\right]\\
	+ \left(\pb -\pb'\right)\cdot\mathbf{v}_\chi\frac{\Enup{'}}{p'}\frac{\partial\delta \! \left(\Enup{}- \Enup{'}\right)}{\partial p'} \left[\fz{'}-\fz{}\right] \bigg]
\end{multline}

We define the polar axis to be in the same direction as the DM velocity, i.e.
\begin{equation*}
	\mu = \hat{\mathbf{v}}_\chi \cdot \hat{\mathbf{p}} \qquad \text{and} \qquad \mu' = \hat{\mathbf{v}}_\chi \cdot \hat{\mathbf{p}}'.
\end{equation*}

We assume the matrix element to have the same form as the one used for Thompson scattering,
\begin{equation}
	\left|M\right|^2 = 6\pi\sigma_0 m_\chi^2 \left(1+\cos^2\theta\right),
\end{equation}
where
\begin{equation*}
	\cos\theta = \hat{\mathbf{p}} \cdot \hat{\mathbf{p}}'~,
\end{equation*}
and $\sigma_0$ is some constant.

The angular part of the collision integral can then be expressed in terms of $\mu'$ and $\phi$,

\begin{multline}
C(p) = \frac{6\pi \left(2\pi\right)^4 \sigma_0 n_\chi}{\left(2\pi\right)^3 \left(2\pi\right)^3 8E_\nu \! \left(\mathbf{p}\right)}
\int \frac{dp' p'^2}{\Enup{'}} \bigg[\\
\delta \! \left(\Enup{} - \Enup{'}\right) \int_{-1}^{1}d\mu'\left[\fo{'}-\fo{}\right]\int_{0}^{2\pi}d\phi' \left(1+\left( \hat{\mathbf{p}} \cdot \hat{\mathbf{p}}'\right)^2\right)\\
+ \frac{\Enup{'}v_\chi}{p'}\frac{\partial\delta \! \left(\Enup{}- \Enup{'}\right)}{\partial p'} \left[\fz{'}-\fz{}\right] \int_{-1}^{1}d\mu'\left[\mu p - \mu'p'\right]\int_{0}^{2\pi}d\phi' \left(1+\left( \hat{\mathbf{p}}\cdot \hat{\mathbf{p}}'\right)^2\right) \bigg].
\end{multline}
We assume a cosmology with azimuthal symmetry, hence $\fo{'}$ depends only on $\mu'$ and not $\phi$. We solve the $\phi$-integral by rewriting the integrand in terms of Legendre polynomials, just as in reference \cite{Dodelson1995ReionizationBackground},
\begin{equation*}
	+\left(\hat{\mathbf{p}} \cdot \hat{\mathbf{p}}'\right)^2 = \frac{4}{3}\left(1+\frac{1}{2}P_2\left(\hat{\mathbf{p}} \cdot \hat{\mathbf{p}}'\right)\right)
\end{equation*}
then using the addition theorem of the spherical harmonics
\begin{equation*}
	\int_{0}^{2\pi} d\phi P_2\left(\hat{\mathbf{p}} \cdot \hat{\mathbf{p}}'\right) = 2\pi P_2\left(\hat{\mathbf{p}} \cdot \hat{\mathbf{v}}_\chi\right)P_2\left(\hat{\mathbf{p}}' \cdot \hat{\mathbf{v}}_\chi\right) = 2\pi P_2\left(\mu\right) P_2\left(\mu'\right)
\end{equation*}
We can also use the definition of the moments of the distribution function,
\begin{equation}
	f_l \left(p\right) = \int_{-1}^{1} \frac{d\mu}{2}P_l\left(\mu\right) f\left(p,\mu\right).
\end{equation}
This allows us to do the $\mu'$ and $\phi$ integrals, resulting in
\begin{multline}
	C(p) = \frac{\pi\left(2\pi\right)^4 \sigma_0 n_\chi 4\pi}{\left(2\pi\right)^3 \left(2\pi\right)^3 E_\nu \! \left(\mathbf{p}\right)}
	\int \frac{dp' p'^2}{E_\nu \! \left(p'\right)} \bigg[\\
	\delta \! \left( E_\nu \! \left(p\right) - E_\nu \! \left(p'\right)\right) \left(
	f^{\left(1\right)}_0 \!\left(p'\right) 
	+ \frac{1}{2}f^{\left(1\right)}_2 \!\left(p'\right) P_2 \! \left(\mu\right) 
	-f^{\left(1\right)} \!\left(p,\mu\right) 
	\right)\\
	+ \frac{ E_\nu \! \left(p'\right)v_\chi p \mu}{p'}\frac{\partial\delta \! \left(E_\nu \! \left(p\right)- E_\nu \! \left(p'\right)\right)}{\partial p'} \left[\fz{'}-\fz{}\right]  \bigg].
\end{multline}
The first part of the integral is easily done using the delta function,
\begin{multline}
	\int \frac{dp' p'^2}{\Enup{'}} \delta \! \left(\Enup{} - \Enup{'}\right) \left(
	f^{\left(1\right)}_0 \!\left(p'\right) 
	+ \frac{1}{2}f^{\left(1\right)}_2 \!\left(p'\right) P_2 \! \left(\mu\right) 
	-f^{\left(1\right)} \!\left(p,\mu\right) 
	\right)\\
	= \frac{p^2}{\Enup{}} \left(
	f^{\left(1\right)}_0 \!\left(p\right) 
	+ \frac{1}{2}f^{\left(1\right)}_2 \!\left(p\right) P_2 \! \left(\mu\right) 
	-f^{\left(1\right)} \!\left(p,\mu\right) 
	\right).
\end{multline}
The second part of the integral is done with integration by parts
\begin{equation}
    \begin{split}
    \int dp' p' v_\chi p \mu \left(\fz{'}-\fz{}\right) &\frac{\partial\delta \! \left(E_\nu \! \left(p\right)- E_\nu \! \left(p'\right)\right)}{\partial p'}\\
	&=v_\chi p \mu \bigg(\left[p' \left(\fz{'}-\fz{}\right) \delta \! \left(E_\nu \! \left(p\right)- E_\nu \! \left(p'\right)\right) \right]_{p'=0}^{p'=\infty}\\
	&\quad-\int dp' \left(\fz{'}-\fz{}\right) \delta \! \left(E_\nu \! \left(p\right)- E_\nu \! \left(p'\right)\right)\\
	&\quad-\int dp' p' \frac{\partial\fz{'}}{\partial p'}\delta \! \left(E_\nu \! \left(p\right)- E_\nu \! \left(p'\right)\right)
	\bigg)\\
	&=-v_\chi p^2 \mu \frac{\partial\fz{}}{\partial p}.
    \end{split}
\end{equation}

Inserting these yields the collision term
\begin{equation}
	C(p) = \frac{\pi\left(2\pi\right)^4 \sigma_0 n_\chi 4\pi}{\left(2\pi\right)^3 \left(2\pi\right)^3 }\left[ \frac{p 2}{E_\nu \! \left(p\right)} \left(
	f^{\left(1\right)}_0 \!\left(p\right) 
	+ \frac{1}{2}f^{\left(1\right)}_2 \!\left(p\right) P_2 \! \left(\mu\right) 
	-f^{\left(1\right)} \!\left(p,\mu\right) 
	\right) - \frac{v_\chi p^2 \mu}{E_\nu \! \left(p\right)} \frac{d\fz{}}{d p} \right],
\end{equation}
which reduces to
\begin{equation}
C(p) = \frac{\ \sigma_0 n_\chi p^2}{E_\nu^2 \! \left(p\right)}\left[ 
f^{\left(1\right)}_0 \!\left(p\right) 
+ \frac{1}{2}f^{\left(1\right)}_2 \!\left(p\right) P_2 \! \left(\mu\right) 
-f^{\left(1\right)} \!\left(p,\mu\right) 
 - v_\chi \mu E_\nu \! \left(p\right) \frac{d\fz{}}{d p} \right].
\end{equation}

\subsection{The Boltzmann Hierarchy}
We will now derive the Boltzmann hierarchy for massive neutrinos with DM interactions, proceeding in a fashion similar to reference \cite{MaBertschinger}. Their Boltzmann equation in the synchronous gauge is
\begin{multline}
	\frac{\partial\Psi}{\partial\tau} + i \frac{p}{E_\nu \! (p)}\left(\mathbf{k}\cdot\hat{\mathbf{n}}\right)\Psi + \frac{d\ln\fz{}}{d\ln p} \left[\dot{\eta}-\frac{\dot{h}+6\dot{\eta}}{2}\left(\hat{\mathbf{k}}\cdot\hat{\mathbf{n}}\right)^2\right] = \\
	\frac{a \sigma_0 n_\chi p^2}{E_\nu^2 \! \left(p\right)}\left[ 
	\Psi_0 \!\left(p\right) 
	+ \frac{1}{2}\Psi_2 \!\left(p\right) P_2 \! \left(\mu\right) 
	-\Psi \!\left(p,\mu\right) 
	- \frac{v_\chi \mu E_\nu \! \left(p\right)}{\fz{}} \frac{d\fz{}}{d p} \right],
\end{multline}
with $\hat{\mathbf{n}}$ being the unit vector in the direction of $p$ and the usual definition of $\Psi$,
\begin{equation*}
	f(p) = f^{\left(0\right)}\!(p)\left(1+\Psi\right).
\end{equation*}
We cannot analytically integrate over all momenta, so we write $\Psi$ as a Legendre series,
\begin{equation}
	\Psi\!\left(\mathbf{k},\hat{\mathbf{n}},p,\tau\right) = \sum_{l=0}^{\infty}\left(-i\right)^l \left(2l+1\right) \Psi_l \! \left(\mathbf{k},p,\tau\right) P_l \! \left(\hat{\mathbf{k}}\cdot\hat{\mathbf{n}}\right).
\end{equation}
Because the velocity $\mathbf{v}_\chi$ is irrotational, its Fourier transform is parallel to $\mathbf{k}$, hence $\hat{\mathbf{v}}_\chi\cdot\hat{\mathbf{n}} = \hat{\mathbf{k}} \cdot \hat{\mathbf{n}} = \mu$. For brevity, we define the interaction rate
\begin{equation}
	C_\chi\!(p) \equiv \frac{a \sigma_0 n_\chi p^2}{E_\nu^2 \! \left(p\right)}.
\end{equation}

The zeroth moment of the differential equation is then
\begin{equation}
\begin{split}
	\frac{\partial \Psi_0}{\partial \tau} &= \int_{-1}^{1}\frac{d\mu}{2}\bigg\{\\
	&\qquad\qquad C_\chi\!(p) \left[ 
	\Psi_0 \!\left(p\right) 
	+ \frac{1}{2}\Psi_2 \!\left(p\right) P_2 \! \left(\mu\right) 
	-\Psi \!\left(p,\mu\right) 
	- \frac{v_\chi \mu E_\nu \! \left(p\right)}{\fz{}} \frac{d\fz{}}{d p} \right]\\
	&\qquad\qquad- i \frac{pk}{E_\nu \! (p)}\mu\Psi - \frac{d\ln\fz{}}{d\ln p} \left[\dot{\eta}-\frac{\dot{h}+6\dot{\eta}}{2}\mu^2\right]\bigg\}\\
	&=-\frac{pk}{\Ep}\Psi_1\!(p)+ \frac{1}{6}\dot{h} \frac{d\ln\fz{}}{d\ln p},
\end{split}
\end{equation}
the same as for the noninteracting case. The parts of the integrals not associated with the interaction will also always yield the same as the standard case, so there is no reason to calculate them again. The first moment is then
\begin{equation}
\begin{split}
\frac{\partial \Psi_1}{\partial \tau} &= \frac{pk}{3\Ep}\left(\Psi_0(p)-2\Psi_2(p)\right) \\
&\qquad+\int_{-1}^{1}\frac{d\mu}{2}C_\chi\!(p) \left[ 
\Psi_0 \!\left(p\right) \mu
+ \frac{1}{2}\Psi_2 \!\left(p\right) P_2 \! \left(\mu\right) \mu 
-\Psi \!\left(p,\mu\right) \mu
- \frac{v_\chi \mu^2 E_\nu \! \left(p\right)}{\fz{}} \frac{d\fz{}}{d p} \right]\\
&=\frac{1}{3}\frac{pk}{\Ep}  \left(\Psi_0(p)-2\Psi_2(p)\right) - C_\chi\!(p)\frac{v_\chi E_\nu \! \left(p\right)}{3\fz{}} \frac{d\fz{}}{d p} - C_\chi\!(p)\Psi_1(p).
\end{split}
\end{equation}
Following the same method for the second moment, we get
\begin{multline}
	\frac{\partial\Psi_2}{\partial\tau} = \frac{1}{5}\frac{pk}{\Ep}  \left(2\Psi_1(p)-3\Psi_3(p\right)
	 - \left(\frac{\dot{h}}{15} + \frac{2\dot{\eta}}{5}\right)\frac{d \ln \fz{}}{d \ln p} - \frac{9}{10}C_\chi\!(p)\Psi_2(p).
\end{multline}
The $C_\chi\Psi$-term will just become a $C_\chi\Psi_l$-term for higher $l$, acting as a damping term. The form of the higher-$l$ moments is thus
\begin{equation}
	\frac{\partial\Psi_l}{\partial\tau} =
	\frac{1}{2l+1}\frac{pk}{\Ep} 
	\left(l\Psi_{l-1}-(l+1)\Psi_{l+1}\right)-C_\chi(p)\Psi_l (p)
\end{equation}